\documentclass{aa}  
\usepackage[varg]{txfonts}
\usepackage{graphicx}

\usepackage{soul}
\usepackage{epstopdf}
\usepackage{longtable}
\usepackage{pdflscape}
\usepackage{graphicx}
\usepackage{hyperref}
\usepackage{amssymb}    % Extra maths symbols
\usepackage{amsmath}
\usepackage{pdflscape}
\usepackage{footnote}

\newcommand{\angstrom}{\text{\normalfont\AA}}
\newcommand{\HII}{H~\textsc{ii}}

\newcommand{\HeI}{{\ion{He}{i}}}
\newcommand{\HeII}{{\ion{He}{ii}}}

\newcommand{\NIII}{{\ion{N}{iii}}}
\newcommand{\CIII}{{\ion{C}{iii}}}
\newcommand{\CIV}{{\ion{C}{iv}}}
\newcommand{\HI}{{\ion{H}{i}}}

\newcommand{\rion}[2]{{\ensuremath{\mbox{\mathrm #1$\,${\small\uppercase\expandafter{\romannumeral#2\relax}}}}}}

\newcommand{\elec}{\ensuremath{{\mathrm e}^{-}}}

\usepackage{newtxtext,newtxmath}
\usepackage[T1]{fontenc}
\usepackage{ae,aecompl}

\usepackage{color}
\usepackage[normalem]{ulem}

\definecolor{darkgreen}{rgb}{0.13, 0.55, 0.13}

\definecolor{brown}{rgb}{0.59, 0.29, 0.0}
\definecolor{ab}{rgb}{0.36, 0.54, 0.66}

\newcommand{\aref}[1]{\hyperref[#1]{Appendix~\ref{#1}}}

\makeatletter
\renewcommand*\aa@pageof{, page \thepage{} of \pageref*{LastPage}}
\makeatother

\let\linenumbers\relax

\begin{document}

   %\title{Empirical and theoretical estimates of HeII ionization budgets}
   
   \title{Strong nebular HeII emission induced by He$^+$ ionizing photons escaping through the clumpy winds of massive stars}
   % \titlerunning{Strong nebular HeII emission}
   %\subtitle{Census of flickering sources}

   \author{A.\,Roy\inst{\ref{inst:sns},\ref{inst:anu}}\thanks{\email{arpita.roy1016@gmail.com}}
        \and M.\,R.\,Krumholz\inst{\ref{inst:anu},\ref{inst:astro3d}} 
\and S. \,Salvadori\inst{\ref{inst:uflorence},\ref{inst:inaf}}
\and G. \,Meynet\inst{\ref{inst:geneva}, \ref{inst:grav}}
\and S. \,Ekstr\"{o}m\inst{\ref{inst:geneva}, \ref{inst:grav}}
\and J. \,S.\, Vink\inst{\ref{inst:armagh}}
\and A.\,A.\,C.\, Sander\inst{\ref{inst:heidel}}
\and R. S. \,Sutherland\inst{\ref{inst:anu},\ref{inst:astro3d}}
\and S. \,Paul\inst{\ref{inst:manchester}, \ref{inst:mcgill}, \ref{inst:capetown}}
\and A. \,Pallottini\inst{\ref{inst:sns}}
\and \'{A}. \,Sk\'{u}lad\'{o}ttir\inst{\ref{inst:uflorence},\ref{inst:inaf}}
          }

   \institute{Scuola Normale Superiore, Piazza dei Cavalieri 7, I-56126 Pisa, Italy \label{inst:sns}
   \and Research School of Astronomy and Astrophysics, Australian National University, Cotter Road, Weston Creek, ACT 2611, Australia \label{inst:anu}
\and ARC Centre of Excellence for All Sky Astrophysics in 3 Dimensions (ASTRO 3D), Canberra, ACT 2611, Australia \label{inst:astro3d}
\and Dipartimento di Fisica e Astronomia, Universit\'{a} degli Studi di Firenze, via G. Sansone 1, 50019, Sesto Fiorentino, Italy \label{inst:uflorence}
\and INAF - Osservatorio Astrofisico di Arcetri, Largo E. Fermi 5, I-50125, Firenze, Italy \label{inst:inaf}
\and D\'{e}partement d'Astronomie, Universit\'{e} de Gen\`{e}ve, Chemin Pegasi 51, CH-1290 Versoix, Switzerland \label{inst:geneva}
\and Gravitational Wave Science Center (GWSC), Universit\'{e} de Gen\`{e}ve, CH-1211 Geneva, Switzerland \label{inst:grav}
\and Armagh Observatory and Planetarium, Armagh, Northern Ireland, BT61 9DG \label{inst:armagh}
\and Zentrum f\"{u}r Astronomie der Universit\"{a}t Heidelberg, Astronomisches Rechen-Institut, M\"{o}nchhofstr. 12-14, 69120 Heidelberg \label{inst:heidel}
\and Jodrell Bank Centre for Astrophysics, School of Physics and Astronomy, The University of Manchester, Manchester M13 9PL, UK \label{inst:manchester}
\and Department of Physics, McGill University, Montreal, QC, Canada H3A 2T8 \label{inst:mcgill}
\and Department of Physics and Astronomy, University of the Western Cape, Robert Sobukhwe Road, Bellville, 7535, South Africa \label{inst:capetown}
}

   \date{Received XXX / Accepted YYY}

% \abstract{}{}{}{}{} 
% 5 {} token are mandatory
 
  \abstract
   {The origin of nebular \HeII~emission in both local and high-redshift galaxies remains an unsolved problem. Various theories have been proposed to explain it, including \HeII-ionizing photons produced by high mass X-ray binaries, ultra-luminous X-ray sources, or ``stripped” He stars produced by binary interaction or evolution of rapidly rotating ($v/v_{\rm{crit}}\gg 0.4$) single massive stars, shock ionization, and hidden active galactic nuclei. All of these theories have shortcomings, however, leaving the cause of nebular \HeII\, emission unclear.}
   {We investigate the hypothesis that the photons responsible for driving nebular \HeII\, emission are produced by the evolution of single massive stars and/or Wolf-Rayet (WR) stars whose winds are on the verge of becoming optically thin due to clumping, thus allowing significant escape of hard ionizing photons. We combined models of stellar evolution with population synthesis and nebular models to identify the most favorable scenarios for producing nebular \HeII~via this channel.}
   {We used the Modules for Experiments in Stellar Astrophysics (\textsc{mesa}) code to compute evolutionary tracks for stars with initial masses of $10\--150\, \mathrm{M}_\odot$ and a range of initial metallicities and rotation rates. We then combined these tracks with a range of custom treatments of stellar atmospheres, which were intended to capture the effects of clumping, in the population synthesis code Stochastically Lighting Up Galaxies (\textsc{slug}) in order to produce the total ionizing photon budgets and spectra. We used these spectra as inputs to \textsc{cloudy} calculations of nebular emission at a range of nebular densities and metallicities.}
   {We find that if WR winds are clumpy enough to become close to optically thin, stellar populations with a wide range of metallicities and rotation rates can produce \HeII\, ionizing photons at rates sufficient to explain the observed nebular $I(\HeII)/I(\mathrm{H}\beta)$ ratio $\sim 0.004\--0.07$ found in \HeII-emitting galaxies. Metal-poor rapidly rotating stellar populations ($[\mathrm{Fe}/\mathrm{H}]=-2.0$, $v/v_\mathrm{crit} = 0.4$) also reach these levels of \HeII\, production, even for partially clumpy winds. These scenarios also yield \HeII, H$\beta$, and ``blue bump'' line equivalent widths comparable to those observed in \HeII\, emitters. Only for homogeneous non-clumpy winds did we fail to find combinations of metallicity and stellar rotation rate that yield $I(\HeII)/I(\mathrm{H}\beta)$ values as high as those observed in \HeII emitters.}
   {Contrary to previous findings, we conclude that single WR stars can be a strong source for nebular \HeII\, emission if their winds are sufficiently clumpy. This scenario also reproduces a range of other properties found in \HeII\, emitters, suggesting that hard photons escaping through clumpy WR winds are a strong candidate to explain nebular \HeII-emission.
   }

               
   \titlerunning{Nebular \HeII\, emission from optically thin WR winds}
   \maketitle

\section{Introduction}\label{sec:intro}
The \HeII $\, \lambda4686$ and \HeII $\,  \lambda1640$ emission lines that are commonly observed in the optical and UV spectra of active galactic nuclei (AGNs) and star-forming (SF) galaxies result from the recombination of He$^{\mathrm{++}}$ with \elec. Production of these lines therefore requires the presence of hard ionizing photons at energies $\geq 54.4$ eV, or wavelengths $\lambda \leq 228\,$ \AA, corresponding to the ionization potential of singly ionized He. The presence of such hard photons is expected in galaxies that host an AGN. However, both direct observations and indirect inferences from population studies have indicated that SF galaxies also manage to produce photons at these energies. Directly, nebular \HeII \, emission has been detected in SF galaxies without any AGN signatures, particularly at subsolar metallicities for local galaxies, and in galaxies at redshifts $z \gtrsim 2$ \citep{shirazi2012, james2022}. Indirectly, AGN number density drops sharply at redshifts $z \gtrsim 3$, while \HeII \, emitters are found up to $z \sim 7$ \citep{cassata2013, stark2016, patricio2016, berg2018, sobral2019, nanayakkara2019, saxena2020a}. \HeII \, emitters indeed become increasingly prevalent at higher redshifts: \citet{kehrig2011, kehrig2016} found that $\gtrsim 3 \%$ of the galaxy population at $z\gtrsim 3$ exhibits narrow \HeII \, lines, which is well above the fraction ($\sim 1\-- 1.5\%$) at $z\sim 0$ \citep{saxena2020a}. The combination of a drop in AGN number density with an increase in \HeII \, emitter fraction at high redshift means that AGNs are unlikely to be significant contributors to \HeII \ emission at $z \gtrsim 3$ (\citealt{feltre2016} and references therein).

While this evidence strongly suggests that SF galaxies can in at least some circumstances produce photons well past the He$^+$ ionization edge, the precise origin of \HeII \, emission in SF galaxies is poorly understood. One possible route is through the production of compact objects at the ends of massive stars' lives. In support of this hypothesis, many local ($z\sim 0$) galaxies exhibiting robust \HeII \, $\lambda4686$ emission are characterized by low metallicities, as exemplified by I Zw18, SBS0335-052E with a metallicity of $Z\sim 1/30 , Z_\odot$ \citep{kehrig2016, kehrig2018}, or the young massive cluster NGC 346 (\citealt{sixtos2023} and references therein) in the Small Magellanic Cloud. This is a significant clue because X-ray luminosity per unit star-formation rate ($L_{\mathrm{X}}$/SFR) and \HeII $\, \lambda4686$ intensity display a strikingly similar anti-correlation with metallicity, which led \citet{schaerer2019} to suggest that high-mass X-ray binaries (HMXB) and/or ultra-luminous X-ray (ULX) sources serve as the primary contributors to nebular \HeII \, emission. \citeauthor{schaerer2019}'s hypothesis is also supported by the fact that HMXBs  are observed to be produced more readily in metal-poor stellar populations. Consistent with this scenario, \citet{mayya2023} found that of the 32 \HeII $\, \lambda4686$ emitting \HII -regions in the Cartwheel galaxy (which has an Large Magellanic Cloud-like metallicity), ten coincide with ULX sources.

However, \citet{senchyna2020} argue that HMXBs by themselves are insufficiently luminous enough to explain observed \HeII \, line strengths and suggest as an alternative modifications to stellar wind prescriptions (which affect massive stellar hard ionizing photon budgets) and/or additional softer X-ray sources alongside HMXBs. In support of this argument, \citet{saxena2020b} found no significant difference between the X-ray luminosities and $L_\mathrm{X}$/SFR ratios of 18 \HeII\, $\lambda$1640-emitting SF galaxies selected from the VANDELS survey at $z\approx 2-4$ and a control sample of non-\HeII~emitters at similar redshifts. Based on their analysis, they concluded that HMXBs and/or any weak or obscured AGNs might not be the dominant \HeII\, producers at $z\gtrsim 2$.

Several alternative physical mechanisms have also been proposed by various authors. Ionization by strong shocks \citep{dopita1996, izotov2012} is one possibility, though \citet{thuan2005} and \citet{izotov2012} argued that this mechanism is viable only if shocks are so strong that they also supply $\sim10 \%$ of the total number of hydrogen ionizing photons produced in a galaxy. However, there is currently no model establishing a quantitative connection between shock strengths and other galaxy properties nor is it clear how a shock-driven model could explain the anti-correlation between \HeII\, emissivity and metallicity in nearby SF galaxies. Another alternative route is via the production of stars with higher effective temperatures and thus harder ionizing spectra than are found in standard stellar evolution models. Such stars could potentially be produced via rotation-induced chemical mixing \citep{szecsi2015, roy2020}, by the rejuvenation of old stars through binary interactions \citep{eldridge2017, gotberg2018, gotberg2019}, or as a late stage in the evolution of metal-free population III stars \citep{tumlinson2000, schaerer2003, visbal2015}. However, more recent quantitative explorations of the idea that ``rejuvenated" stars produced by binary interactions explain \HeII\, emission have found that they are insufficient to reproduce the observed \HeII \, intensity in subsolar local galaxies \citep{stanway2019}.

Wolf-Rayet (WR) stars represent yet another possible route to \HeII\, emission. The WR phase in a massive star's life begins, in general, during the core He burning phase \citep[e.g.,][]{roy2020, roy2021a}, and during this phase, stars reach the high effective temperatures, $T_{\mathrm{eff}} \sim 80 \--100$ kK, required to produce photons capable of ionizing He$^+$ \citep{schaerer2002}. As with the HMXB scenario, this idea has received observational support from an observed correlation between \HeII~emission and spectral features, indicating a strong spectral contribution from WR stars. Some nearby SF galaxies emitting \HeII\, $\lambda4686$ display a ``blue bump'' in their spectra, a distinctive signature of the presence of WR stars that is associated with the \HeII\, $\lambda4686$ line blended with either \NIII \, $\lambda4640$ in WN stars and/or \CIII /\CIV \, $\lambda4650$ in WC stars \citep{guseva2000, lopez2010, kehrig2016, mayya2020, mayya2023}. \HeII~emitting galaxies also occasionally exhibit a ``red bump'' as well \citep{lopez2010}, which is linked to the broad \CIV\, $\lambda 5808$ line primarily observed in WC stars. The I(\HeII~$\lambda 4686$)/I(H$\beta$) ratio -- onwards abbreviated to $I(\HeII)/I(\mathrm{H}\beta)$ -- in galaxies with sub-solar metallicity that show prominent WR features in their integrated spectra is a few percent \citep{izotov2004}, which is much higher than is typical of galaxies lacking WR features. On smaller scales, \citet{mayya2023} measured $I(\HeII)/I(\mathrm{H}\beta) \sim 0.004 - 0.07$, with a mean of $0.01 \pm 0.003$ for the 32 \HeII \,$\lambda 4686$ emitting \HII \, regions in the Cartwheel galaxy.

However, the WR idea has proven difficult to test for both observational and theoretical reasons. Observationally, the problem has  partly been one of resolution. Most early attempts to explain \HeII~features in spectra have focused on spectra from the Sloan Digital Sky Survey, which typically encompass regions spanning several kiloparsecs around galactic centers. Over such large spatial scales, many physical mechanisms may contribute to \HeII~emission, including photoionization by single and/or binary stellar populations, strong shock-driven ionization, He$^+$ ionization by ``stripped" He stars, HMXBs, and even the ionization by weak AGNs. The poor resolution makes it challenging to disentangle these processes and check stellar population models against observations. For example, a number of authors have found that stellar population models fail to reproduce the observed amount of \HeII~flux in these integrated galaxy spectra \citep[e.g.,][]{plat2019, schaerer2019, stanway2019}, but this may simply be because on kiloparsec scales, the spectra also contain significant contributions from supernova (SN)-driven shock ionization \citep{plat2019}.

Observations focused on individual nebulae at a few hundred parsec scales represent a more promising way of testing WR models (and other models where \HeII~is driven partly by young stellar populations). However, even this approach has proven challenging. For example, even with $\sim 500$ pc observations of the metal-poor ($Z \sim 0.03\, Z_\odot$) nearby galaxy SBS 0335-052E using optical VLT/MUSE spectroscopic and Chandra X-ray observations, \citet{kehrig2018} failed to reach any solid conclusion as to the origin of \HeII~emission. Their findings only indicated that the observed \HeII \, ionization budget could be produced either by single rotating metal-free ($Z \lesssim 10^{-5} \, Z_\odot$) stars or by a metal-poor ($Z\sim 0.01 \, Z_\odot$) binary population with a top-heavy initial mass function (IMF). \citet{kehrig2015} concluded that WR stars are insufficient to explain the rate of He$^+$ ionization demanded by their observations of another metal-poor nearby galaxy, IZw 18. By contrast, \citet{mayya2020} carried out MEGARA observations of the central starburst cluster in NGC 1569 and concluded that WR stars can produce enough He$^+$ ionizing photons to explain the observed \HeII \, $\lambda 4686$ flux. \citet{mayya2023} reached similar conclusions from $\sim 120$ pc resolution MUSE observations of regions around super star clusters ($M_* \sim 10^5 - 10^7$ M$_\odot$) in the Cartwheel Galaxy.

On the theoretical side, the WR scenario has proven challenging to test because of uncertainties on the He$^+$-ionizing fluxes of WR stars.  The largest uncertainty in these models concerns the atmospheres of these stars, with some theories assuming a spherical, homogeneous, optically thick WR atmosphere. Such models are clearly inconsistent with observations that suggest that winds of most WR stars consist of stochastic clumps \citep{schumann1975, moffat1988, eversberg1998, grosdidier2001, lepine2008, chene2020} rather than having a homogeneous structure. This feature is difficult to capture in simulations of stellar atmospheres because most of the simulations are 1D spherically symmetric. However, there is a standard approximation method to deduce clumping factors assuming that the clumps themselves remain optically thin, known as ``microclumping", from the electron
scattering wings \citep{hillier1991, koesterke1999}. Incorporating microclumping leads to a reduction in the derived mass-loss rates from the observations of WR stars but a constant transformed mass-loss rate.\footnote{The transformed mass-loss rate is a normalized measure to approximately remove the effect of different luminosities and terminal velocities. It can be written as $\dot{M_t} \propto \dot{M} v_\infty D_\infty^{-1/2} L^{3/4}$, where $\dot{M}$ is the mass-loss rate, $v_\infty$ is the terminal velocity, $L$ is the luminosity, and $D_\infty$ is the clumping factor defined as $D_{\infty} = \langle \rho^2\rangle/\langle\rho\rangle^2$, where $\langle\rho\rangle$ is the mean density of the atmosphere.} \citet{sander2023} finds that if this transformed mass-loss rate is below a characteristic value $\log[\dot{M}/(\mathrm{M}_\odot \, \mathrm{yr}^{-1})] \approx -4.5$, WR winds become transparent to He$^+$-ionizing photons, potentially greatly increasing the escape of such photons into the interstellar medium. However, this approximation is not sufficient to calculate the escape of He$^+$-ionizing photons from a stellar wind; such a calculation requires full 3D (or at least 2D) simulations. A few such simulations do exist in the literature: \citet{oskinova2007} carried out Monte Carlo simulations for very approximate treatments of clumpiness in WR stars, and \citet{moens2022} recently presented the first ever hydrodynamic models of WR winds. However, \citeauthor{moens2022}'s simulations are restricted to only low mass WR stars $\sim 10 \, \mathrm{M}_\odot$. Wind porosity is another topic that is not yet explored well. \citet{owocki2004} introduced a new ``porosity-length" formalism in the 1D atmosphere modeling and discussed how wind porosity reduces the effective opacity of the atmosphere, thus boosting the escape of ionizing photons. However, there are no detailed quantitative studies nor any 2D or 3D modeling of wind porosity in the literature. Thus, neither the mass loss of WR winds nor the wind opacity and/or clumpiness and wind porosity are well-quantified.

Our goal in this paper is to revisit the hitherto unsolved problem of nebular \HeII\, emission and demonstrate that incorporating close to optically thin and/or higher clumpiness and porous WR atmosphere models \citep[e.g.,][]{nugis2000, vink2001} can significantly enhance the production of He$^+$ ionizing photons. Models with this new atmosphere treatment can explain both the observed \HeII \, flux as well as the $I(\HeII)/I(\mathrm{H}\beta)$ ratio in \HII \, regions through the single stellar population channel. Importantly, this explanation avoids the need to invoke additional physical mechanisms such as binarity, ``stripped" He stars, shock ionization, HMXBs, and/or contribution from AGNs. In addition to this primary result, we provide the first comprehensive studies exploring the impact of mass, metallicity, and rotation rates of WR stars on the ability to produce ionizing spectra hard enough to explain observed  \HeII~emission.

The remainder of this paper is organized as follows. In Sect. \ref{sec:method_sec}, we describe our method for calculating the expected nebular \HeII\, fluxes from stellar evolution and population synthesis models. In Sect. \ref{sec:result_sec}, we present the results of our population synthesis modeling, focusing on the time evolution of ionizing photon luminosities and their ratios and their dependence on various model parameters.We also analyze the nebular lines these photons drive, calculating intensity ratios and equivalent widths (EWs) of key emission lines so that we can compare our models to observations. We conclude the paper by summarizing our findings and their implications in Sect. \ref{sec:conclusion_sec}.

\section{Methods}\label{sec:method_sec}
Our objective is to obtain He$^+$ ionizing photon budgets produced by the single massive star channel, and then use these to compute nebular fluxes in the  \HeII $\, \lambda4686$ and \HeII $\,  \lambda1640$ emission lines. For this purpose we first produce a set of stellar tracks and atmospheres for single massive stars that have high $T_{\mathrm{eff}}\sim 80 \-- 100$ kK (Sect. \ref{sec:stelmod_sec}) and then process them through a stellar population synthesis code (Sect. \ref{sec:popmod_sec}) to obtain the ionizing photon budgets and spectra expected from stellar populations. Finally, we process the star cluster spectra through a photoionization code to derive the nebular spectra (Sect. \ref{sec:specmod_sec}). 

\subsection{Stellar models}\label{sec:stelmod_sec}
%As we are interested in \HeII \, emission via WR pathways, we require a set of stellar tracks of massive stars that are capable to produce He$^+$ ionizing photons. 
Massive stars on the main sequence (MS) lose mass via winds with high velocity ($\sim \mathrm{a\, few}\times 1000$ km/s) and low mass-loss rates $\dot{M} \sim 10^{-6} \, \mathrm{M}_\odot \, \mathrm{yr}^{-1}$, yielding optically thin atmospheres. However, when these stars transition to the WR phase, in general during core He burning, their mass-loss rates typically increase by an order of magnitude to $\dot{M} \sim 10^{-5} \, \mathrm{M}_\odot \, \mathrm{yr}^{-1}$, resulting in an extended optically thick atmosphere. Despite their high opacity, these atmospheres often exhibit clumpiness, which potentially allows ionizing photons to escape through the atmosphere and into the interstellar medium through optically thin pathways. Our goal here is to parameterize this effect.

We begin from a set of stellar tracks, for which purpose we use \textsc{mesa} Isochrone Stellar Tracks (MIST; \citealt{choi2016}) for two rotation rates, $v/v_{\mathrm{crit}} =0.0$, $0.4$ and for three metallicities, [Fe/H] $= 0.0$, $-1.0$, $-2.0$ with a upper mass limit of $150 \, \mathrm{M}_\odot$. At high masses the tracks have a mass interval $\Delta M = 10$ M$_\odot$, and we show in \aref{app:massres} that this relatively high resolution (compared to some other sets of tracks) is important to capture ionizing photon budgets accurately. For details of the MIST setup, such as adopted physical processes, parameter choices, etc., we refer the readers to \citet{choi2016}. These tracks yield, for each initial stellar mass, rotation rate, and metallicity, a set of stellar parameters as a function of time, the most important of which for what follows are the mass-loss rate $\dot{M}$, stellar radius $R$, stellar luminosity $L$, and surface abundances of H, He, C, N, and O; we use these abundances to classify stars as either non-WR or WR, and within WR as one of several possible sub-types: WNL, WNE, WO, WC; see \citealt{choi2016} (who in turn take their scheme from \citealt{georgy2012a}) for full details of how the classification is performed. For further detailed discussions on WR subtypes, we refer the readers to \citealt{roy2020, roy2021a}.

While we use the MIST tracks as a default because they are well-tested, they have a weakness that, like most stellar tracks, they assume that all elements follow solar-scaled abundances. However, it is well established that at metallicities of 1/10th solar and below this assumption is invalid. Instead, $\alpha$-elements are enhanced due to the contributions from type-II supernovae without corresponding type-Ia events (\citet{nicholls2017}, and references therein). To ensure that our results do not depend on the assumption of solar-scaled abundances, in \autoref{app:trackcomp} we repeated the analysis of the main text using the \textsc{stromlo stellar tracks (SST)}\footnote{\href{https://sites.google.com/view/stromlotracks/home?authuser=0}{https://sites.google.com/view/stromlotracks/home?authuser=0}} \citep{grasha2021}, which use empirically motived ``Galactic Concordance'' abundances from \citeauthor{nicholls2017} rather than solar-scaled abundances. 

In order to predict the stellar spectrum at each point along an evolutionary track (see Sect. \ref{sec:popmod_sec}), we require two more pieces of information: the effective temperature and radius of the stellar photosphere, $T_\mathrm{eff}$ and $R_\mathrm{eff}$. For stars without optically thick winds, these are simple: the photosphere is at the stellar surface, $R_\mathrm{eff} = R$, and the effective temperature is then given implicitly by the usual relation $L = 4\pi R_\mathrm{eff}^2 \sigma_\mathrm{SB} T_\mathrm{eff}^4$. For WR stars with potentially opaque winds, however, the choice is much subtler because in such stars, the photosphere could potentially be at a much larger radius than that from which the wind is launched, leading to a much lower effective temperature. We therefore consider three possible approaches to calculating $R_\mathrm{eff}$ and $T_\mathrm{eff}$ for WR stars.

\subsubsection{Highly clumped and porous winds}
\label{sssec:highly_clumped_winds}
One option is to assume that the wind is sufficiently clumped and porous that the solid angle over which it is optically thick covers only a minority of the stellar surface, in which case the effects of the wind are small and we can set $R_\mathrm{eff}$ and $T_\mathrm{eff}$ for WR stars exactly as for all other stars. This is the limit of close to optically thin winds, and is the default option for the MIST tracks, i.e., the values of $R_\mathrm{eff}$ and $T_\mathrm{eff}$ reported by \citet{choi2016} are computed in this way. Note that these stars observationally might not actually be WR stars given their low emission line strengths due to more transparent winds, even though their surface abundances and $T_{\rm{eff}}$ ranges match with WR classifications. These stars rather fall in the category of envelope-stripped stars with a more transparent wind. We consider this wind treatment case just to achieve the maximum limit possible for \HeII~emission from the single massive stars.

\subsubsection{Spherically homogeneous smooth winds}
\label{sssec:spherical_winds}

A second possibility is to treat the wind as spherical and homogenous, and compute the location of the photosphere under this assumption. As a proxy for the cooler temperatures resulting from these extended atmospheres of stars with strong mass loss, we adopt the prescription of \citet{schaller1992}, which we briefly summarize here for readers' convenience. This scheme assumes the wind follows the standard velocity law
\begin{equation}
v(r) = v_{\infty} \left( 1- \frac{R}{r} \right)^\beta \, ,
\label{eq:vel}
\end{equation}
where $v(r)$ is the radial velocity, $\beta$ is the acceleration index for the wind, and $v_{\infty} $ and $R_\mathrm{eff}$ are the terminal velocity and stellar radius, respectively. Following \citet{pauldrach1986}, we adopt $\beta = 2$ and $v_\infty = 2\times 10^3$ km s$^{-1}$. To compute the optical depth through this wind, we note that the flux-weighted mean opacity can be written as 
\begin{equation}
\bar{\kappa} = \sigma_{\mathrm{e}} \left [ 1+ M \right ] \, , 
\label{eq:opacity}
\end{equation}
where $\sigma_{\mathrm{e}}$ is the opacity for electron scattering and $M$ is the force-multiplier denoting the radiative acceleration due to lines in units of $g_{\mathrm{Eddington}}$ ($=\kappa L/ 4 \pi c r^2$, where $c$ is the speed of light, and $\kappa$ is the wind opacity). The term $\sigma_{\mathrm{e}}$ can be written as (Eq. $8.93$ of \citealt{lamers1999})
\begin{equation}
\sigma_{\mathrm{e}} = \frac{\sigma_{\mathrm{T}} n_{\mathrm{e}}}{\rho} = 0.401 \left(I_{\mathrm{H}}X + I_{\mathrm{He}}\frac{Y}{4} + I_{\mathrm{Z}}\frac{Z}{14}\right)\,\mathrm{cm}^2\,\mathrm{g}^{-1},
\label{eq:sigma_e}
\end{equation}  
where $\sigma_{\mathrm{T}} = 6.6524 \times 10^{-25} \, \mathrm{cm}^2$ is the Thomson scattering cross section of electrons, $\rho$ is the total mass density, $n_{\mathrm{e}}$ is the electron density, $I_{\mathrm{H}}=1$, $I_{\mathrm{He}}=2$ and $I_{\mathrm{Z}}=14$ are the number of electrons per ion of H, He and heavier elements, respectively, and $X$, $Y$, and $Z=1-X-Y$ are the mass-fractions of H, He, and heavier atoms at the stellar surface (which we know from the stellar tracks). Therefore, the optical depth in terms of $\bar{\kappa}$ is 
\begin{equation}
\tau(r) = \int_r^{\infty} \bar{\kappa} \rho\, \mathrm{dr} = \int_r^\infty \sigma_{\mathrm{e}} \rho \,\mathrm{dr} + \int_r^\infty \sigma_{\mathrm{e}} M \rho \, \mathrm{dr} = \tau_{\mathrm{e}} +\tau_{\mathrm{lines}} \, ,
\label{eq:tau}
\end{equation} 
where
\begin{equation}
\tau_{\mathrm{e}} = { {\sigma_{\mathrm{e}} \dot{M}} \over {4 \pi v_{\infty}(1-\beta) R} } \left[ 1 - { {1} \over {(1-R/r)^{\beta -1}} } \right] \, 
\label{eq:tau_e}
\end{equation} 
is the electron scattering optical depth and $\rho = \dot{M} / 4\pi r^2 v(r)$ is the mass density in the wind for a star with mass-loss rate $\dot{M}$. We calculated $\tau_{\mathrm{lines}}$ from the prescription of \citet{kudritzki1989} for the force-multiplier $M$ as
\begin{equation}
M \left (\rho, r, v, \frac{dv}{dr}, n_{\mathrm{e}} \right ) = k \left( { {\sigma_{\mathrm{e}} \rho v_{\mathrm{th}}} \over {dv/dr} } \right)^{-\alpha} \left( { {n_{\mathrm{e}}} \over {W(r)} } \right)^\delta \mathrm{CF} \left( r, v, \frac{dv}{dr} \right)   \, ,
\label{eq:M}
\end{equation}
where $v_{\mathrm{th}} = \sqrt{{ {2k_{\mathrm{B}}T_{\mathrm{eff}}} / {m_{\mathrm{H}}} }}$ is the thermal velocity of protons, $k_{\mathrm{B}}$ and $m_{\mathrm{H}}$ are the Boltzmann constant and proton mass respectively, $W$ and $\mathrm{CF}$ are the dilution and correction factors respectively coming from the non-radial streaming of photons, and $k$, $\alpha$, and $\delta$ are numerical fitting factors that are obtained by comparing the analytic prescription above to numerical results; \citeauthor{kudritzki1989} follow \citet{pauldrach1986} in setting $k = 0.124$, $\alpha = 0.64$, and $\delta = 0.07$, and we do the same. We then computed $n_\mathrm{e} / W(r)$ as \citep[Eq. 42]{kudritzki1989}
\begin{eqnarray}
\left(\frac{n_\mathrm{e}}{W(r)} \right)^\delta & = & \Delta 2^\delta \left[q(\delta, \beta)\left(\frac{R}{r}\right)^2 + 1 \right] \\
\Delta & = & \left( { {\dot{M}} \over {4 \pi R^2 v_\infty} } {2 \over {m_{\mathrm{H}}} } { {1+I_{\mathrm{He}}Y} \over {1+4Y} } \times 10^{-11} \, \mathrm{cm}^3 \right ) ^ \delta \,  
\label{eq:Del} 
\end{eqnarray}
and the numerical factor $q(\delta,\beta) = 0.244$ for our chosen values of $\delta$ and $\beta$. Similarly, we computed $\mathrm{CF}$ as \citep[Eq. 15]{kudritzki1989}
\begin{equation}
\mathrm{CF} = { 1 \over {\alpha+1}} { {x^2} \over {1-h(x, \beta)} } \left[ 1 - \left( 1 - {1 \over {x^2}} + h(x, \beta) {1 \over {x^2}} \right) ^{\alpha+1} \right] \, ,
\label{eq:cf}   
\end{equation} 
where $x\equiv r/R$ and $h(x,\beta) \equiv (x-1)/\beta$.

The expressions above fully specify the force multiplier $M$ at each radius, and therefore put us in a position to compute $\tau(r)$ for arbitrary $r$ by evaluating the integrals in Eq. \ref{eq:tau}. We can in turn then find the radius $r$ for which $\tau(r) = 2/3$, which defines our effective radius $R_\mathrm{eff}$ and implicitly our effective temperature $T_\mathrm{eff}$. We carry out this calculation for every point in our stellar evolution calculation where stars are classified as WR. We consider this extreme case of maximum wind opacity just to get a hard lower limit for the \HeII~emission via the single massive star channel, similar to how we use the another extreme scenario for the highly clumped and porous winds (Sect. \ref{sssec:highly_clumped_winds}) to achieve the upper limit.

\subsubsection{Partially clumped winds}

The previous two schemes represent the limits of very strongly clumped and porous winds, and completely smooth winds, but we can also consider an intermediate case motivated by the results of three-dimensional simulations. \citet{moens2022} performed the first ever time-dependent 3D radiation-hydrodynamical simulations of WR star winds, and tabulate the effective temperatures of the resulting structures (their Table 2). Comparing their results for a high-luminosity $10 \, \mathrm{M}_\odot$ star with a clumping factor of $\approx 2$ to the $T_\mathrm{eff}$ we compute for a star of comparable mass and luminosity using the spherically homogeneous wind prescription described in Sect. \ref{sssec:spherical_winds}, we find that the 3D simulation yields a factor of $\approx 2$ larger $T_\mathrm{eff}$. As a crude estimate of the effects of partial clumping, we took $T_\mathrm{eff}$ for this case to be twice the value returned using the spherical wind prescription from Sect. \ref{sssec:spherical_winds}. We note that this is a conservative lower limit for the effects of partial clumping, since $10 \, \mathrm{M}_\odot$ is at the low mass end for WR stars and 2 is a relatively low clumping factor, and we expect the difference between the smooth and clumped value of $T_\mathrm{eff}$ to be larger for higher stellar masses and wind clumping factors.

We caution that this approximation is rather crude, since in reality if the wind has partial coverage of the stellar surface then a better description than a single intermediate effective temperature might be a weighted sum of the outputs of atmospheres with two different effective temperatures and surface gravities -- one representing the contribution from parts of the stellar surface that are directly visible from infinity and one representing areas where the wind is opaque enough to obscure the surface and the photosphere from much further out. Given the non-linear dependence of hard ionizing photon production on $T_\mathrm{eff}$ (as we shall see below), the results from such a mixing model might be quite different from the results we obtain here by specifying a single, intermediate $T_\mathrm{eff}$. However, the development of such improved models for wind partial covering is beyond the scope of this paper, and we therefore resorted to the single $T_\mathrm{eff}$ approximation.

\subsection{Population synthesis modeling}\label{sec:popmod_sec} 

Having obtained values for $R_{\mathrm{eff}}$ and $T_{\mathrm{eff}}$ at each point in the stellar tracks via the three possible methods described above, we next generate synthetic spectra for stellar populations by using these stellar tracks in the Stochastically Lighting Up Galaxies (\textsc{slug}\footnote{\href{https://bitbucket.org/krumholz/slug2/src/master/}{https://bitbucket.org/krumholz/slug2/src/master/}}) code \citep{da-silva2012a, krumholz2015b}. We compute spectra for star clusters with ages from $0.1-10$ Myr with a time step of 0.1 Myr. Given that the observed clusters that produce \HeII~emission are likely substantially more massive than $\sim 10^3-10^4$ M$_\odot$, the range below which stochastic sampling generally becomes important, we turn off \textsc{slug}'s stochastic features and treat the IMF as fully sampled; for this mode the total cluster mass simply acts as a multiplicative constant on the output spectrum. We assume a Chabrier IMF \citep{chabrier2003}. We generate spectra using \textsc{slug}'s ``starburst99'' model for stellar atmospheres, which replicates the model used in starburst99 \citep{leitherer1999, vazquez2005b}. This model uses a library of Kurucz model atmospheres taken from \citet{lejeune1997a} for stars without strong winds, models from \citet{pauldrach2001} for OB stars with optically thin winds, and models from \citet{hillier1998a} for WR stars.

\subsection{Nebular modeling: Spectral synthesis models}\label{sec:specmod_sec}

To generate nebular spectra, we use the spectra produced by \textsc{slug} as inputs to a calculation of the structure and emission from a photoionized nebula using \textsc{cloudy}\footnote{\href{https://gitlab.nublado.org/cloudy/cloudy}{https://gitlab.nublado.org/cloudy/cloudy}} \citep{chatzikos2023}. To perform these calculations we normalize the total luminosity of the stellar population to that of a cluster of mass $10^6$ M$_\odot$, and place an innermost zone for the nebula at a distance of 1 pc from this population. We then use \textsc{cloudy} to calculate a hydrostatic structure for the resulting nebula, adopting \textsc{cloudy}'s ``expanding sphere'' prescription for H~\textsc{ii} region geometry and setting the nebula to be isobaric at a pressure $\log(P/k_{\rm{B}}\, [\mathrm{K}\, \mathrm{cm}^{-3}]) = 7$. We choose our mass normalization and pressure to be typical of values inferred for observed \HeII-emitting regions, but we have verified that varying these choices over plausible ranges leads to negligible differences in the qualitative results for the line ratios in which we are interested below.\footnote{The sole exception to this statement is for the case of a stellar population with $v/v_\mathrm{crit} = 0.0$, $\mathrm{[Fe/H]} = -1$ for the highly clumped and porous winds at 5 Myr, for which at the chosen pressure and luminosity the system admits two distinct solutions for the structure of the nebula, one lower temperature and higher density and one lower density and higher temperature, with somewhat different line ratios; changing the cluster mass or pressure can then cause one of the two solution branches to disappear. However, since this is only one case out of a large number of models, we do not worry about this complication further.} We set the abundances of the 30 non-hydrogenic elements included in \textsc{cloudy} to values consistent with those assumed in the MIST models from which we generate the stellar tracks; this means that nebular abundances properly scale with stellar ones. We further adopt \textsc{cloudy}'s default grain composition for H~\textsc{ii} regions of solar metallicity, scaling the total grain abundance by factors of $0.1$ and $0.001$ for [Fe/H] $=-1$ and $-2$ respectively, following \citet{remy2014}'s findings with regard to variation of the dust-to-gas ratio with metallicity (see the discussion in Fig. 4 of \citealt{remy2014}). We stop running the models when either the electron temperature falls below $1000$ K or the electron fraction, defined as the ratio of electron to total hydrogen densities, falls below $0.1$. In addition to stellar radiation, we also include \textsc{cloudy}'s default cosmic ray background.

\section{Results}\label{sec:result_sec}
Here we describe the results of our nebular \HeII \, modeling. We start in Sect. \ref{ssec:ion_sec} by examining ionizing photon budgets and their dependence on model parameters, most importantly on our adopted treatment of stellar wind clumping. In Sect. \ref{ssec:syn_spec_sec} we analyze synthetic spectra of \HII\, regions, testing under what circumstances our models can reproduce key features of observed \HeII~emitters such as \HeII \, to H$\beta$ intensity ratios and the equivalent widths of prominent lines such as H$\beta$, \HeII \,, and the blue bump lines.

\subsection{Ionizing photon budgets}\label{ssec:ion_sec}
In this section we quantify how these various choices in our treatment of stellar evolution and atmosphere modeling affect ionizing photon budgets.  

\subsubsection{Extended atmosphere correction}\label{sssec::atm_sec}

%%%%%%%%%%%%%%%%%%%%%%%%%%%%%%%%%%%%%%%%%%%%%%%%%%%%%%%%%%%%%%%%%%%%%%%%%%%%%%%%%%%%%%%
\begin{figure}
   \centering
   \includegraphics[width=1.0\columnwidth]{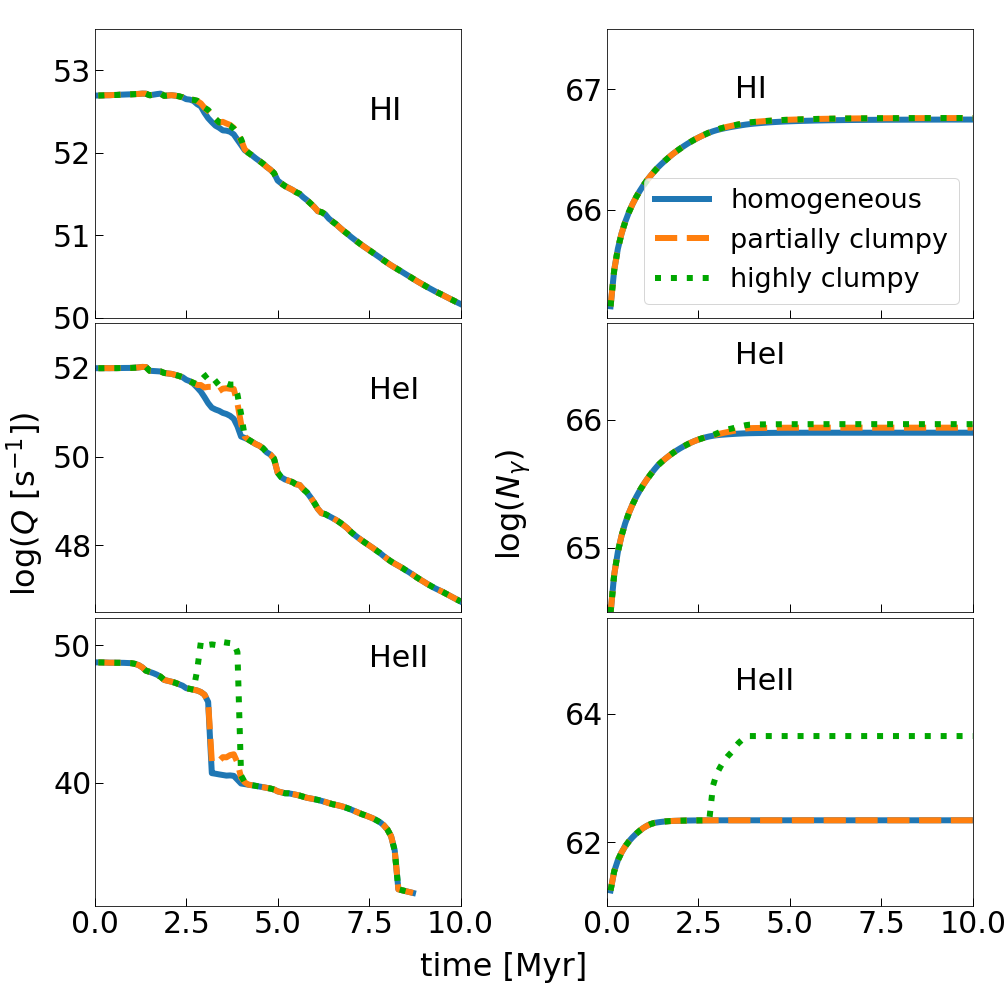}
      \caption{Left: Time evolution of \HI \,, \HeI\,, \HeII\, photon luminosities $Q(t)$, number of photons emitted per unit time, for the three wind prescriptions: optically thick and homogeneous winds, partially clumpy, and highly clumped close to optically thin winds. The results shown are for non-rotating solar metallicity stars and for a star-cluster mass of $10^6\, \mathrm{M}_\odot$. Right: Cumulative photon emission, $N_\gamma = \int_0^T Q(t)\, dt$, as a function of time $T$ for the same parameters.
            }
         \label{fig:atm_comp}
\end{figure}
%%%%%%%%%%%%%%%%%%%%%%%%%%%%%%%%%%%%%%%%%%%%%%%%%%%%%%%%%%%%%%%%%%%%%%%%%%%%%%%%%%%%%%%

We start our inspection with the most crucial yet most uncertain parameter: the scheme used to model clumpy WR atmospheres, as described in Sect. \ref{sec:stelmod_sec}. In Fig. \ref{fig:atm_comp} we show the time evolution of the instantaneous \HI-, \HeI-, \HeII-ionizing photon luminosities (left), along with the cumulative output in these three bands (right), for our three wind opacity schemes: spherically homogeneous optically thick winds, partially clumpy, and very clumpy and porous (close to optically thin winds). This plot is for the case $[\mathrm{Fe}/\mathrm{H}]=0.0$, $v/v_{\rm{crit}}=0.0$ and for a cluster mass of $M_{\rm{cl}}=10^6\, \mathrm{M}_\odot$; we explore the effects of sub-solar metallicity and rotation below. We find that \HI\, photon luminosities have similar values irrespective of wind opacities, with the homogeneous and highly clumpy winds' results differing by only $\sim 3\%$ once they reach the asymptotic plateau at $\gtrsim 5$ Myr. The differences are much larger for harder photons: cumulative photon output differs by $\sim 17\%$ for \HeI, and $\sim 2000\%$ (a factor of 20) for \HeII.

%%%%%%%%%%%%%%%%%%%%%%%%%%%%%%%%%%%%%%%%%%%%%%%%%%%%%%%%%%%%%%%%%%%%%%%%%%%%%%%%%%%%%%%
\begin{figure}
   \centering
   \includegraphics[width=0.95\columnwidth]{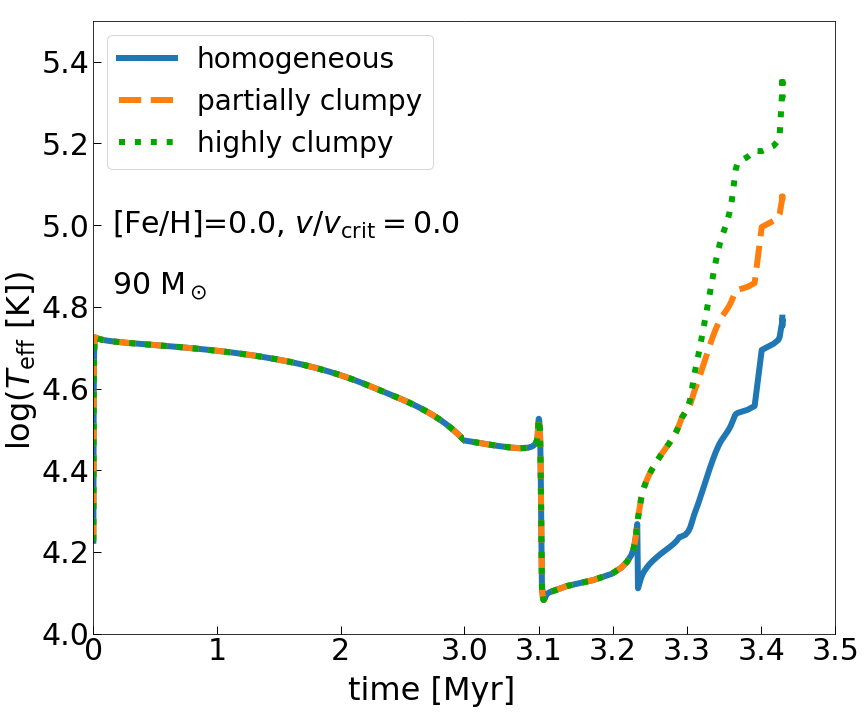}
      \caption{$T_{\rm{eff}}$ versus time for $90 \, \mathrm{M}_\odot$ stars with $[\mathrm{Fe}/\mathrm{H}]=0.0$ and $v/v_{\rm{crit}}=0.0$, with WR winds modeled with three different atmosphere morphologies. We divided the time axis into two equal halves, one from $0\-- 3$ Myr and another from $3\-- 3.5$ Myr, using different stretches in order to zoom-in on the evolution of $T_{\rm{eff}}$ after 3 Myr.}
         \label{fig:teff}
\end{figure}
%%%%%%%%%%%%%%%%%%%%%%%%%%%%%%%%%%%%%%%%%%%%%%%%%%%%%%%%%%%%%%%%%%%%%%%%%%%%%%%%%%%%%%%

The difference between the models is almost entirely due to the rate of hard ionizing photon production at times from $\approx 2.5 - 4$ Myr, when stars $\gtrsim 70$ M$_\odot$ enter their WR phases. To understand the origin of this difference in more detail, we focus on one example case: a non-rotating, solar metallicity star with an initial mass of 90 M$_\odot$. We examine the time evolution of $T_{\rm{eff}}$ for this star in Fig. \ref{fig:teff}, which shows that the optically thick atmosphere reaches a maximum temperature of $\log(T_{\rm{eff}}) \approx 4.8$ during the WR phase, too cool to produce significant quantities of harder photons capable of ionizing \HeII~and with limited production of \HeI-ionizing photons. In contrast, the partially clumpy wind achieves a temperature as high as $\log(T_{\rm{eff}}) = 5.0$, but not higher, leading to the substantial production of \HeI\, photons, but still few harder photons past the \HeII~ionization edge. For the highly clumped, close to optically thin atmosphere, the surface temperature reaches $\log(T_{\rm{eff}}) \gtrsim 5.2$ at $\gtrsim 3$ Myr, resulting in the significant production of photons capable of ionizing both \HeI\, and \HeII. Given this large difference, for the remainder of this paper we focus on highly clumped winds as our fiducial case unless explicitly stated otherwise, with the intention of exploring the maximum \HeII \, ionizing photon budgets and the corresponding line intensities achievable via single WR-star pathways.

\subsubsection{Rotation}\label{sssec:vvcrit_sec}
%%%%%%%%%%%%%%%%%%%%%%%%%%%%%%%%%%%%%%%%%%%%%%%%%%%%%%%%%%%%%%%%%%%%%%%%%%%%%%%%%%%%%%%
\begin{figure}
   \centering
   \includegraphics[width=0.95\columnwidth]{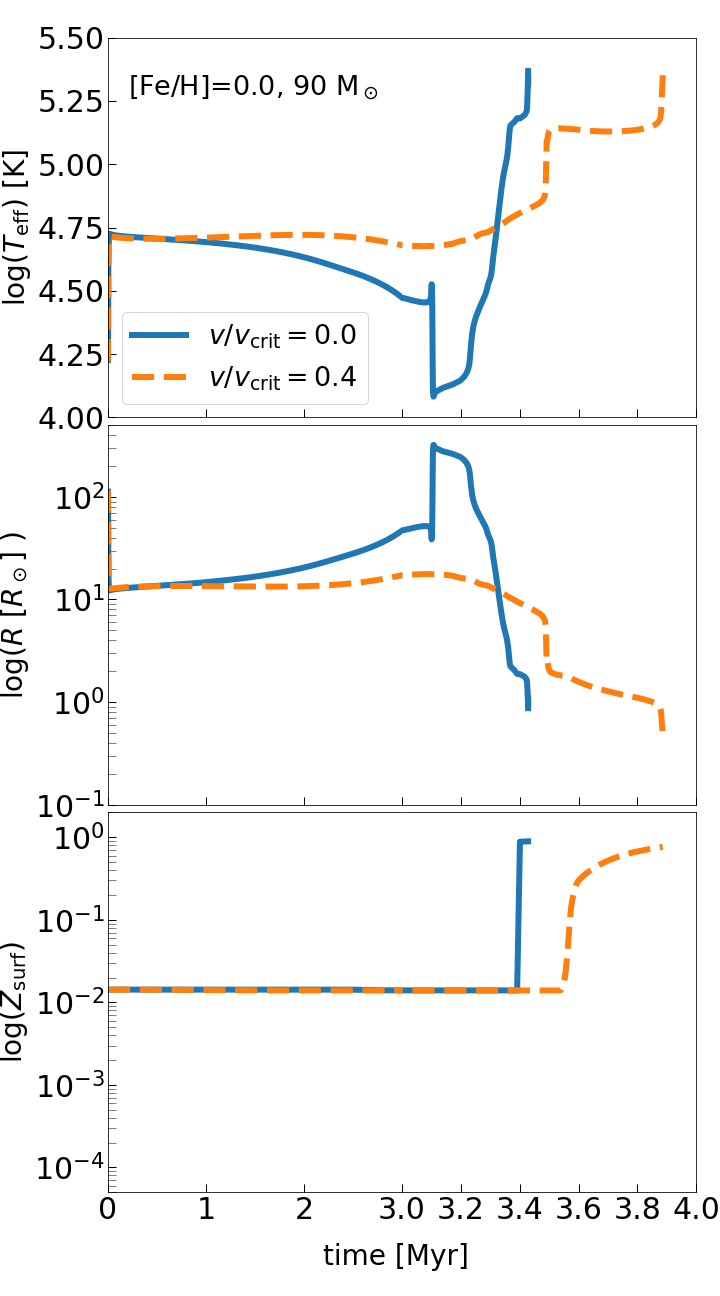}
      \caption{Time evolution of $\log(T_{\rm{eff}})$, top panel; $R$, middle; and surface metal abundance $Z_{\rm{surf}}$, bottom, for $90 \, \mathrm{M}_\odot$ solar-metallicity stars rotating at $v/v_{\rm{crit}} = 0.0$ and $0.4$. We measured these parameters at the surface of the
hydrostatic core because we calculated them for the highly clumpy winds. We divided the time axis into two equal halves as in Fig. \ref{fig:teff}, but here the second half of the range extends from $3\--4$ Myr.} 
         \label{fig:teff_vvcrit_comp}
\end{figure}
%%%%%%%%%%%%%%%%%%%%%%%%%%%%%%%%%%%%%%%%%%%%%%%%%%%%%%%%%%%%%%%%%%%%%%%%%%%%%%%%%%%%%%%
Now that we have established that our models can in some cases produce large quantities of \HeII-ionizing photons when the winds are highly clumped, we next examine the importance of stellar rotation. From the standpoint of production of hard ionizing photons, rotation has two major effects:
\begin{itemize}
\item {Longer MS and WR lifetime}: Rotation enhances the mixing of chemical elements in a massive star, leading to chemically homogeneous evolution (CHE) in the most extreme cases for stars$\gtrsim 90\, \mathrm{M}_\odot$. This increased mixing extends the supply of hydrogen fuel to the core, prolonging the star's MS lifetime. However, following the typical definition of WR stars, having high $T_{\rm{eff}}$ ($\gtrsim 10^4$ K) and high helium mass-fraction ($\gtrsim 40\%$), rotating stars with masses $\gtrsim 70 \, \mathrm{M}_\odot$ enter the WR phase when they are still on the MS (core H burning). Moreover, the WR lifetime also enhances for rotating stars because the helium fuel supply to the core lasts longer in the most extreme cases for stars $\gtrsim 90 \, \mathrm{M}_\odot$, similar to the extended hydrogen fuel supply during the MS. Thus, the combined effect of the early entry to the WR phase and the prolonged lifetime makes the rotating star to spend a longer duration in the WR phase.    
\item {Enhanced surface opacity}: Rotation may enhance the heavy element abundances at the surface of WR stars, and hence in the winds, through two channels. First, rotation directly enhances the mass-loss rate in the MIST framework following the prescription $\dot{M}(v) = \dot{M}(0) \left( 1/ (1-v/v_{\rm{crit}}) \right)^\zeta$, where $\dot{M}(v)$ is the mass-loss rate as a function of rotation velocity $v$, and $\zeta$ is set to $0.43$ based on numerical calibrations (\citealt{friend1986, langer1998}; see Sect. 3.7.3 of \citealt{choi2016} for details). Second, rotation-induced instabilities bring heavy elements from the core to the stellar surface, and this combined with rotation-enhanced mass loss exposing deeper layers in the star yields a significant enhancement in the abundance of heavy elements at the stellar surface \citep{roy2020}. Both the higher mass loss and the larger heavy element abundances increase the wind opacity. 
\end{itemize}

%%%%%%%%%%%%%%%%%%%%%%%%%%%%%%%%%%%%%%%%%%%%%%%%%%%%%%%%%%%%%%%%%%%%%%%%%%%%%%%%%%%%%%%
\begin{figure}
   \centering
   \includegraphics[width=1.0\columnwidth]{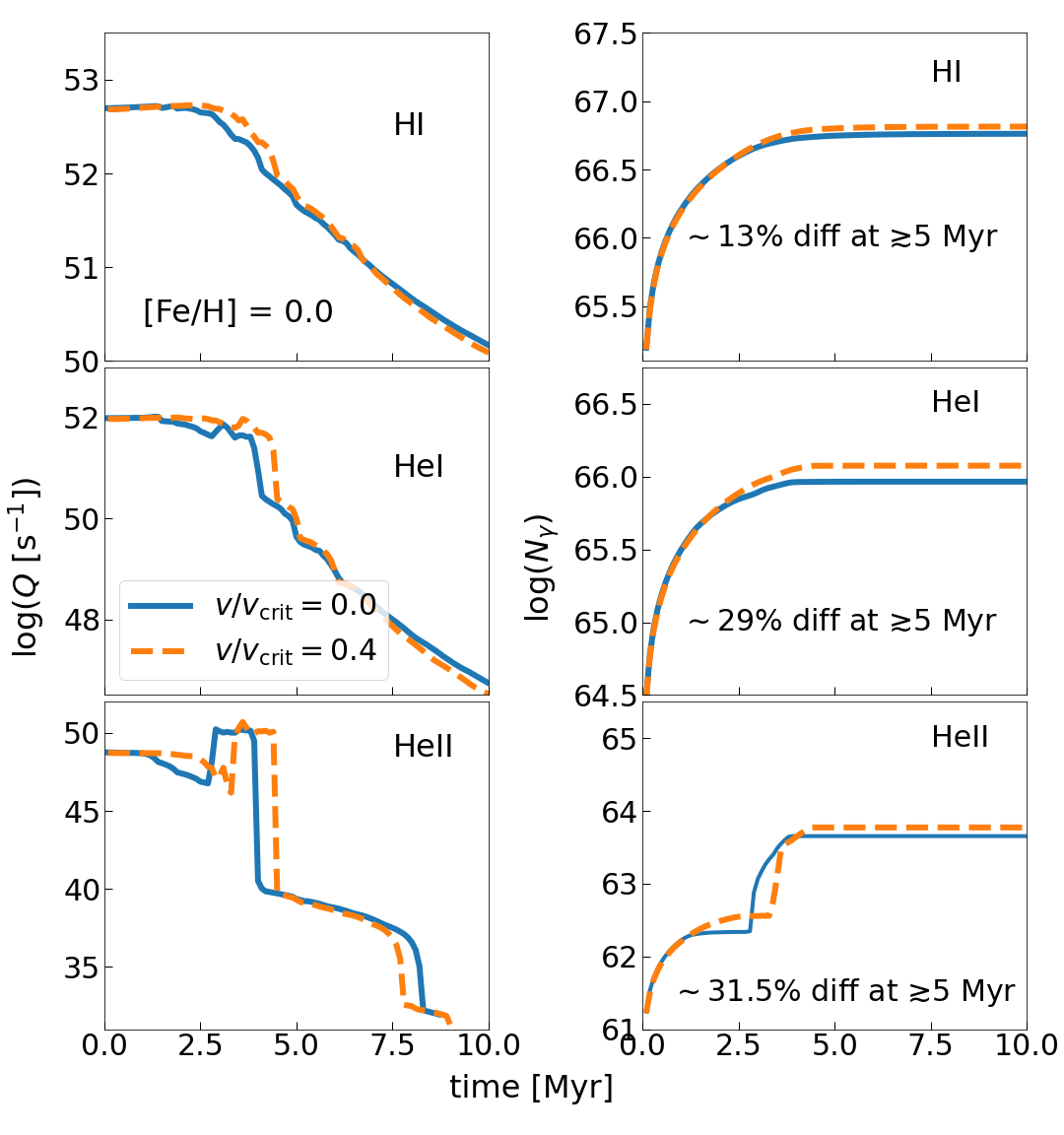}
      \caption{Same as Fig. \ref{fig:atm_comp} but now showing the results for highly clumped and porous winds for two different rotation rates $v/v_{\rm{crit}} = 0.0$ and $0.4$.} 
         \label{fig:vvcrit_comp_feh_0}
\end{figure}
%%%%%%%%%%%%%%%%%%%%%%%%%%%%%%%%%%%%%%%%%%%%%%%%%%%%%%%%%%%%%%%%%%%%%%%%%%%%%%%%%%%%%%%

To explore the effect of rotation, in Fig. \ref{fig:teff_vvcrit_comp} we show the time evolution of $T_{\rm{eff}}$, radius, and the surface metallicity $Z_{\rm{surf}}$ (defined as the mass fraction in elements heavier than He) for the same 90 M$_\odot$ solar-metallicity case shown in Fig. \ref{fig:teff} but now comparing the cases $v/v_{\rm{crit}}=0.0$ and $0.4$; the former is the same as the highly clumpy winds shown in Fig. \ref{fig:teff}. We find that the non-rotating star inflates toward the end of the MS, resulting in decreased $T_{\rm{eff}}$. Subsequently, it begins losing mass rapidly due to the combination of low surface gravity and high mass ($\gtrsim 80 \, \mathrm{M}_\odot$) and thus luminosity. As a result, the star shrinks, exposing the ``fossil" convective core and producing high surface metal abundance and $T_{\rm{eff}}$ soon after the star enters the WR phase. However, this phase is short-lived ($\sim 0.2$ Myr). In contrast, the rotating star spends a longer duration in the WR phase because of the early entry into this phase. As the star enters the WR phase, both $T_{\rm{eff}}$ and $Z_{\rm{surf}}$ increase significantly, while the radius slowly decreases.

Having understood the evolution of a single massive star at two different rotation rates, we now return to the broader question of how rotation affects IMF-averaged hard photon production. In Fig. \ref{fig:vvcrit_comp_feh_0}, we plot the ionizing photon budgets for $10^6$ M$_\odot$, solar-metallicity clusters with highly clumped winds at $v/v_{\rm{crit}} = 0$ and 0.4; the $v/v_\mathrm{crit} = 0$ case shown in this figure is identical to the highly clumped winds shown in Fig. \ref{fig:atm_comp}. We see that the longer duration spent as a WR star in the rotating case leads to higher budgets of \HI-, \HeI-, and \HeII-ionizing photons. Consistent with what we found when examining treatments of clumping, sensitivity is greatest for the highest-energy photons: we find differences of $\sim 15\%$, $\sim 30\%$ and $\sim 32\%$ for the cumulative budgets of \HI, \HeI, and \HeII, respectively, at $\gtrsim 5$ Myr. However, recall that changing the clumping factor produced a factor of $\approx 20$ change in the \HeII-ionizing photon budget; in comparison, we see that the effects of rotation are substantially smaller.

%%%%%%%%%%%%%%%%%%%%%%%%%%%%%%%%%%%%%%%%%%%%%%%%%%%%%%%%%%%%%%%%%%%%%%%%%%%%%%%%%%%%%%%
%\begin{figure}
%   \centering
%   \includegraphics[width=1.0\columnwidth]{vvcrit_comp_feh_m1.png}
%      \caption{Blah blah..}
%         \label{fig:vvcrit_comp_feh_m1}
%\end{figure}
%%%%%%%%%%%%%%%%%%%%%%%%%%%%%%%%%%%%%%%%%%%%%%%%%%%%%%%%%%%%%%%%%%%%%%%%%%%%%%%%%%%%%%%

%%%%%%%%%%%%%%%%%%%%%%%%%%%%%%%%%%%%%%%%%%%%%%%%%%%%%%%%%%%%%%%%%%%%%%%%%%%%%%%%%%%%%%%
%\begin{figure}
%   \centering
%   \includegraphics[width=1.0\columnwidth]{vvcrit_comp_feh_m2.png}
%      \caption{Blah blah..}
%         \label{fig:vvcrit_comp_feh_m2}
%\end{figure}
%%%%%%%%%%%%%%%%%%%%%%%%%%%%%%%%%%%%%%%%%%%%%%%%%%%%%%%%%%%%%%%%%%%%%%%%%%%%%%%%%%%%%%%

\subsubsection{Metallicity}\label{sssec:feh_sec}
%%%%%%%%%%%%%%%%%%%%%%%%%%%%%%%%%%%%%%%%%%%%%%%%%%%%%%%%%%%%%%%%%%%%%%%%%%%%%%%%%%%%%%%
\begin{figure}
   \centering
   \includegraphics[width=1.0\columnwidth]{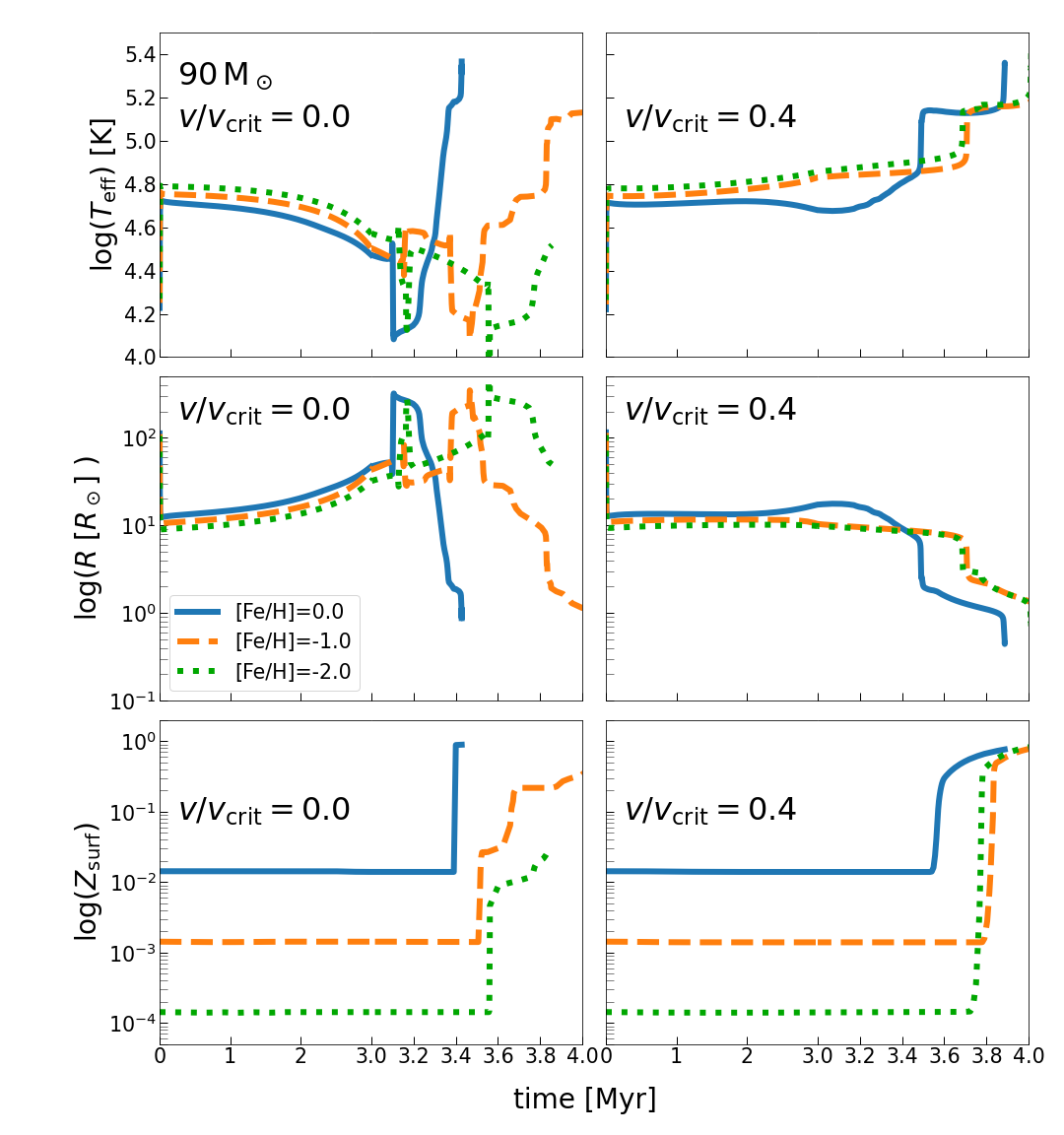}
      \caption{Same as Fig. \ref{fig:teff_vvcrit_comp} but now showing the result for three different initial metallicities: $[\mathrm{Fe}/\mathrm{H}]=0.0$ (blue solid lines), $-1.0$ (orange dashed lines), and $-2.0$ (green dotted lines). The left panels correspond to $v/v_{\rm{crit}} = 0.0$, and the right panels show $v/v_{\rm{crit}}=0.4$.} 
         \label{fig:teff_feh_comp}
\end{figure}
%%%%%%%%%%%%%%%%%%%%%%%%%%%%%%%%%%%%%%%%%%%%%%%%%%%%%%%%%%%%%%%%%%%%%%%%%%%%%%%%%%%%%%%

We next examine the effect of the initial stellar metallicity, which also affects hard ionizing photon production. Surface metallicity is a determining factor for mass loss, with higher metallicities resulting in higher loss rates due to stronger line-driven winds.\footnote{We note that metal line-driven mass-loss rates are actually dependent on the iron surface abundances, not on the total surface metallicity (\citet{sander2020a} and references therein). However, our MIST models account for the surface metallicity in order to keep it similar to the other existing 1D stellar evolution codes (for example, the \href{https://www.unige.ch/sciences/astro/evolution/en/database}{\textsc{Geneva Stellar Tracks}}).} More metal-rich stars therefore have larger mass-loss rates prior to the WR phase, and this higher envelope loss in turn makes it easier to expose stars' ``fossil"-convective cores (cores that are no longer convective, but were part of the convective cores in the previous evolutionary stages; for details, see \citealt{roy2020}), which are heavily helium-enhanced due to CNO burning. This leads to the earlier entry into the WR phase.

%%%%%%%%%%%%%%%%%%%%%%%%%%%%%%%%%%%%%%%%%%%%%%%%%%%%%%%%%%%%%%%%%%%%%%%%%%%%%%%%%%%%%%%
\begin{figure}
   \centering
   \includegraphics[width=1.0\columnwidth]{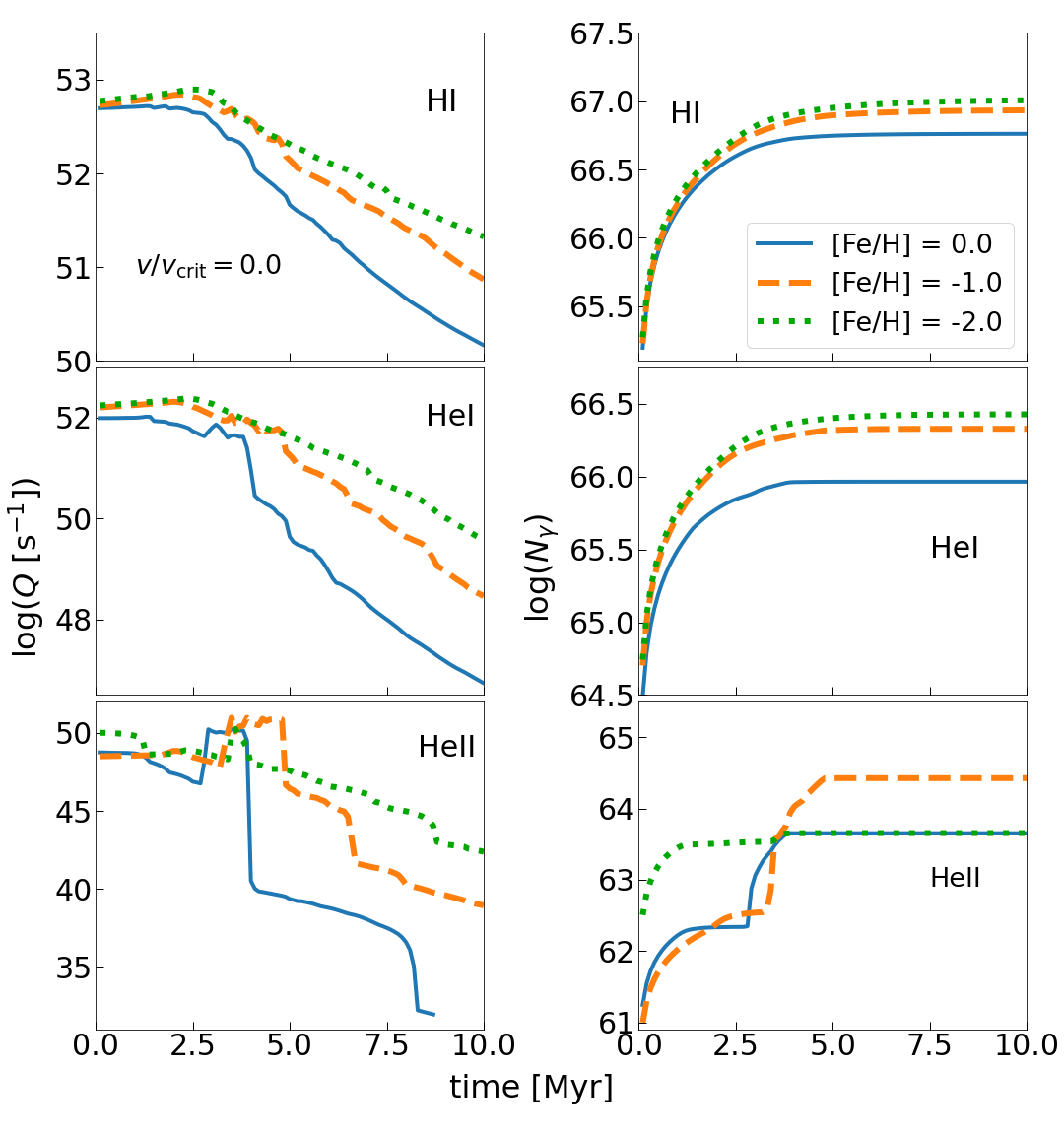}
      \caption{Same as Fig. \ref{fig:atm_comp} but now showing the results for three different metallicities, $[\mathrm{Fe}/\mathrm{H}] = 0.0$, $-1.0$, and $-2.0$, for non-rotating stars. The $[\mathrm{Fe}/\mathrm{H}] = 0.0$ case shown here is identical to the highly clumped winds shown in Fig. \ref{fig:atm_comp}.}
         \label{fig:feh}
\end{figure}
%%%%%%%%%%%%%%%%%%%%%%%%%%%%%%%%%%%%%%%%%%%%%%%%%%%%%%%%%%%%%%%%%%%%%%%%%%%%%%%%%%%%%%%

In Fig. \ref{fig:teff_feh_comp} we show the metallicity dependence through the time evolution of $T_{\rm{eff}}$ (top panels), radius (middle) and surface $Z_\mathrm{surf}$ (bottom panels), again focusing on the example of a $90 \, \mathrm{M}_\odot$ star; we show both the non-rotating (left panels) and rotating, $v/v_{\rm{crit}} = 0.4$ (right panels) cases. For non-rotating stars, we observe that as $[\mathrm{Fe}/\mathrm{H}]$ decreases, the MS lifetime increases significantly: $\sim 3.3$ Myr for solar metallicity versus $\sim 3.7$ Myr for $[\mathrm{Fe}/\mathrm{H}] = -1.0$. This is because at low metallicity, the initial hydrogen mass fraction is larger compared to the metal-rich star, and also the convective core mass increases as the metallicity decreases; the combined effect of which enhances the hydrogen fuel reservoir for the metal-poor star enhancing its MS lifetime. As we approach the WR phase, the stronger mass loss in the two more metal-rich cases ($[\mathrm{Fe}/\mathrm{H}] = 0.0$ and $-1.0$) causes the stars to shrink rapidly and expose their metal-rich fossil convective cores, producing a sudden increase in $T_{\rm{eff}}$ and surface metallicity; this effect is greatly diminished for $[\mathrm{Fe}/\mathrm{H}] = -2.0$. In contrast, with rotation (right panels), the surface parameters become nearly independent of metallicity because rotation-induced mixing leads to CHE for stars $\gtrsim 90 \, \mathrm{M}_\odot$, thus bringing helium from the core to the surface and allowing stars to enter the WR phase even with lower mass losses. Consequently, the MS lifetimes are only weakly dependent on $[\mathrm{Fe}/\mathrm{H}]$, with only about a 3\% difference between solar and $1/10$th solar metallicity and a less than 1\% difference between $[\mathrm{Fe}/\mathrm{H}] = -1.0$ and $-2.0$.

%%%%%%%%%%%%%%%%%%%%%%%%%%%%%%%%%%%%%%%%%%%%%%%%%%%%%%%%%%%%%%%%%%%%%%%%%%%%%%%%%%%%%%%
\begin{figure}
   \centering
   \includegraphics[width=1.0\columnwidth]{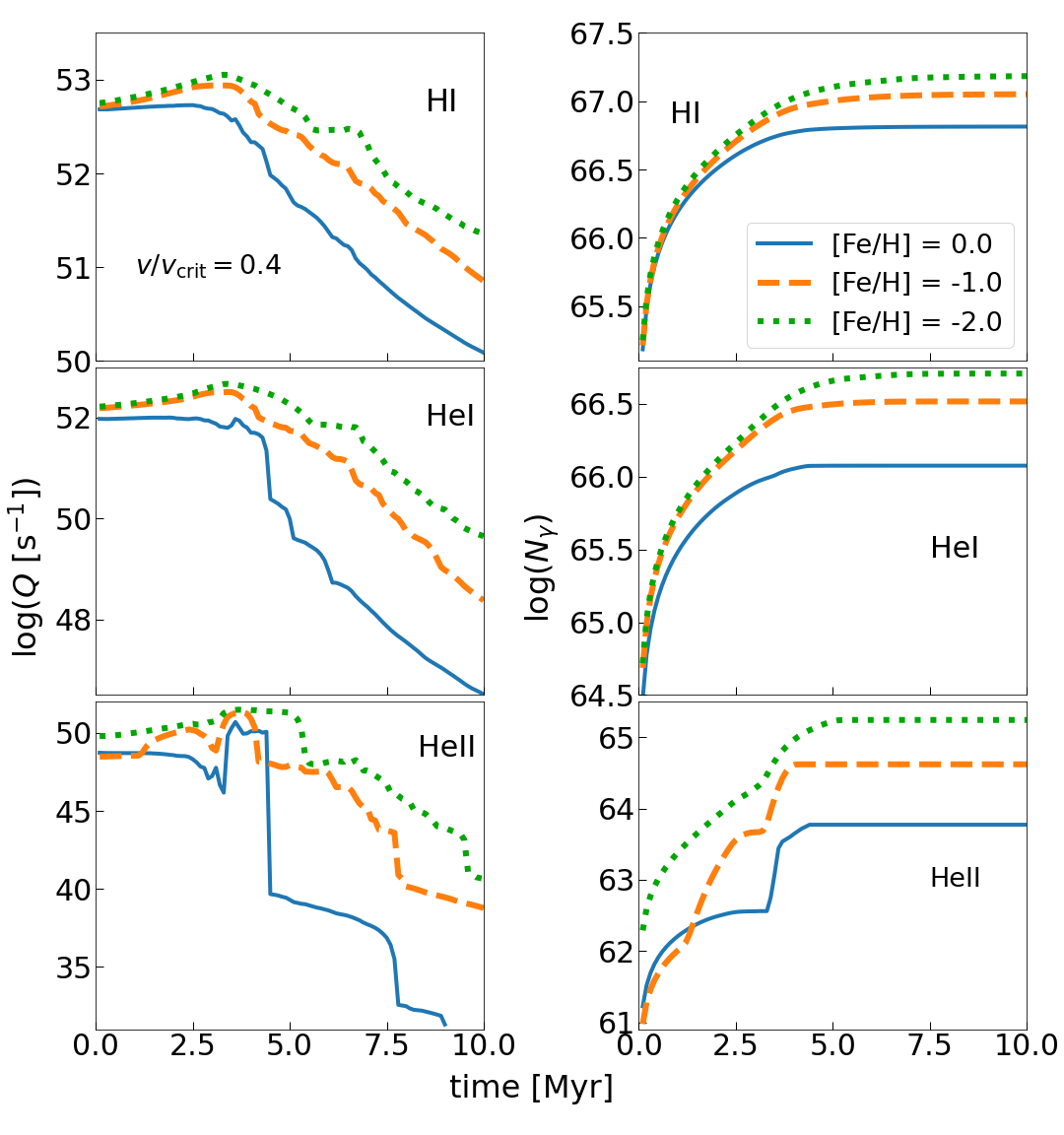}
      \caption{Same as Fig. \ref{fig:feh} except the stars are now rotating at $v/v_{\rm{crit}}=0.4$.}
         \label{fig:feh_vvcrit04}
\end{figure}
%%%%%%%%%%%%%%%%%%%%%%%%%%%%%%%%%%%%%%%%%%%%%%%%%%%%%%%%%%%%%%%%%%%%%%%%%%%%%%%%%%%%%%%

We show how these changes affect the IMF-averaged ionizing photon budgets in Fig. \ref{fig:feh} and Fig. \ref{fig:feh_vvcrit04} for the non-rotating and rotating cases, respectively. We find that for softer ionizing photons (\HI, \HeI) in non-rotating stars, the ionizing luminosity increases monotonically as $[\mathrm{Fe}/\mathrm{H}]$ decreases due to decreasing surface opacity and thus decreasing wind opacity and increasing surface temperature. For \HeII, on other hand, we find that ionizing photon output is maximized at $[\mathrm{Fe}/\mathrm{H}]=-1.0$, and is lower at both $-2.0$ and $0$. This pattern holds true only for non-rotating stars. We can understand this non-monotonic behavior as a result of a competition between two effects. One is the same mechanism we have already invoked to explain the softer photons: as we lower the metallicity we lower the opacity, raising the surface temperature and increasing the escape of ionizing radiation. The other competing effect is mass loss: stars lose more mass at higher metallicity, which in turn means that once stars enter the WR phase, they have smaller radii and higher effective temperatures, favoring production of harder ionizing photons as the metallicity increases. The competition between these effects yields a maximum for \HeII\ ionizing photon production at $[\mathrm{Fe}/\mathrm{H}]=-1.0$. The effect has importance only for the non-rotating case because this is the case for which the initial surface metallicity has the impact on stellar mass loss. For this case, $[\mathrm{Fe}/\mathrm{H}]=-1.0$ yields a cumulative \HeII-ionizing photon production that is a factor of $\approx 5$ larger than either the metallicity 0 or $-2$.

On the contrary, in rapidly rotating stars, CHE for the most massive stars $\gtrsim 90 \, \mathrm{M}_\odot$ facilitates the early entry to the WR phase prolonging the WR lifetime, even at low metallicities, despite reduced metal line-driven mass loss, resulting in nearly uniform surface abundances across different metallicities. Consequently, there is a monotonicity in the increment in ionizing luminosity as $[\mathrm{Fe}/\mathrm{H}]$ decreases for both softer and harder ionizing photons. We find that for this case with $v/v_{\rm{crit}} = 0.4$, $[\mathrm{Fe}/\mathrm{H}]=-2.0$ yields a cumulative \HeII-ionizing photon budget that is a factor of $\approx 29$ and $\approx 4$ larger than that of solar metallicity and $-1$, respectively. 
To conclude, rotating stars with $v/v_{\rm{crit}}\gtrsim 0.4$ provide a more favorable scenario for the \HeII\, ionizing photon production across all metallicities.

\subsubsection{From ionizing photon budgets to line ratios}\label{sssec:qratio_sec}
%%%%%%%%%%%%%%%%%%%%%%%%%%%%%%%%%%%%%%%%%%%%%%%%%%%%%%%%%%%%%%%%%%%%%%%%%%%%%%%%%%%%%%%
\begin{figure}
   \centering
   \includegraphics[width=1.0\columnwidth]{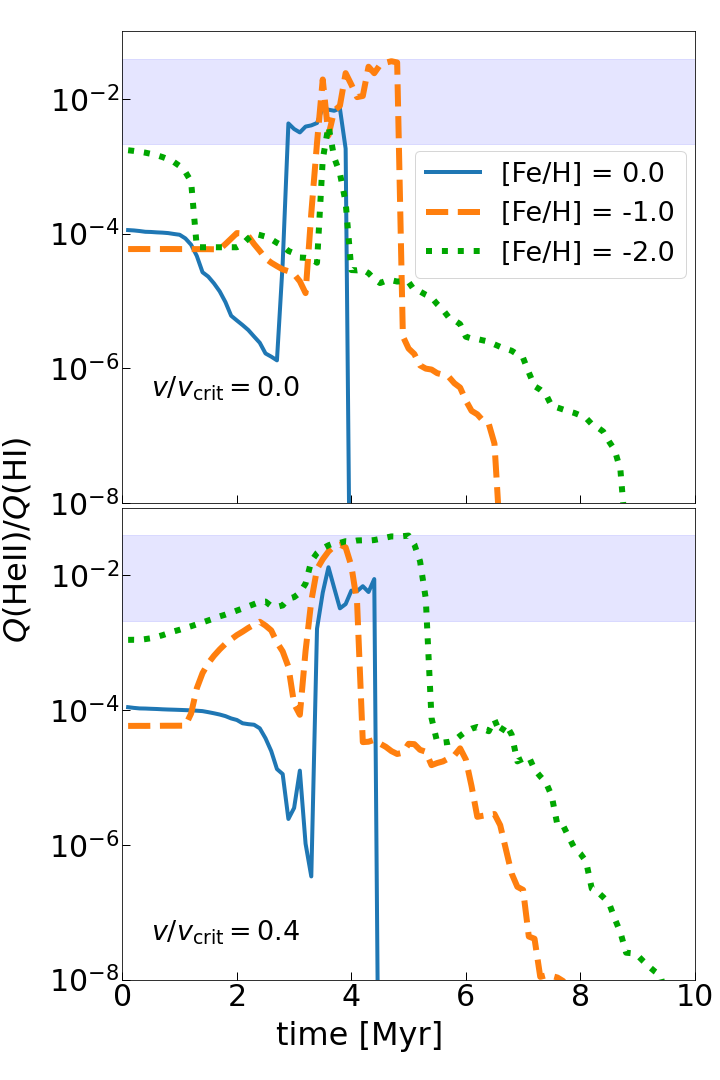}
      \caption{Time evolution of $Q(\HeII)/Q(\HI)$ for non-rotating (top panel) and rotating stars with $v/v_{\rm{crit}}=0.4$ (bottom panel) for three metallicities, $[\mathrm{Fe}/\mathrm{H}] = 0.0$ (blue solid line), $-1.0$ (orange dashed), $-2.0$ (green dotted), and for highly clumped winds. The blue band shows the approximate range of $Q(\HeII)/Q(\HI) \sim 0.0021\-- 0.0371$ expected to produce line intensity ratios $I(\HeII \, \lambda4686)/I(\mathrm{H}\beta)$ in the range observed in spatially resolved observations of the Cartwheel Galaxy \citep{mayya2023}, $\approx 0.004\-- 0.07$.}
         \label{fig:qhe2_qh1}
\end{figure}
%%%%%%%%%%%%%%%%%%%%%%%%%%%%%%%%%%%%%%%%%%%%%%%%%%%%%%%%%%%%%%%%%%%%%%%%%%%%%%%%%%%%%%%

Having discussed the dependence of various stellar parameters on the ionizing photon budgets, we are now in a position to address the central, motivating question for this study, namely, what are the circumstances that allow stars with clumpy winds to drive the bright nebular \HeII~lines seen in some SF galaxies. In Sect. \ref{ssec:syn_spec_sec}, we carry out full nebular photoionization calculations.

Our simple approximation is to treat emission in both the \HeII $\, \lambda4686$ and H$\beta$ lines as arising purely from recombinations occurring in a uniform-temperature, dust-free, photon-bounded \HII~region within which all photons above the \HeII~ionization threshold are absorbed by He$^+$ ions and all photons below that energy but above 1 Ryd are absorbed by neutral H. Since the number of photons above the \HeII~threshold is small compared to the total ionizing flux, in this case we can simply equate the \HI-ionizing flux with the rate of recombinations to \HI, with a fixed fraction of those recombinations yielding emission of a photon in the H$\beta$ line, and we obtained the usual relation between ionizing photon emission and H$\beta$ line luminosity:
\begin{equation}
L(\mathrm{H}\beta) = E_{\mathrm{H}\beta} \alpha_B(\mathrm{H}\beta, T) Q(\HI) \approx 1.94\times 10^{-13} Q(\HI)\,\mathrm{erg},
\end{equation}
where $\alpha_B(\mathrm{H}\beta, T)$ is the effective recombination coefficient for H$\beta$ emission in case B as a function of the temperature $T$, and our numerical evaluation is appropriate for $T \approx 10^4$ K. Similarly, we can equate the recombination rate to \HeII~with the production rate of \HeII-ionizing photons and compute the luminosity of the \HeII $\, \lambda4686$ line as \citep{kehrig2015}
\begin{equation}
L(\HeII) = E_{\HeII} \alpha_B(\HeII , T) Q(\HI) \approx 3.66\times 10^{-13} Q(\HeII)\,\mathrm{erg},
\end{equation}
where $E_{\HeII}$ and $\alpha_B(\HeII, T)$ are the \HeII $\, \lambda4686$ photon energy and recombination rate coefficient. Combining these results, we find that under our simple approximations the intensity ratio for the \HeII $\, \lambda4686$ and H$\beta$ lines should be related to the emission rates of \HI- and \HeII-ionizing photons by
\begin{equation}
\frac{I(\HeII)}{I(\mathrm{H}\beta)} = 1.89 \frac{Q(\HeII)}{Q(\HI)}.
\end{equation}
\citet{kehrig2015} show that varying the temperature leads to only small variations in this ratio.

Since typical observed $I(\HeII)/I({\rm{H}}\beta)$ ratios in \HeII-emitting nebulae such as those found in the Cartwheel Galaxy are $\sim 0.004 \-- 0.07$ \citep{mayya2023}, this simple analysis suggests that stellar populations that produce ionizing photon ratios $Q(\HeII)/Q(\HI)\sim 0.0021 \-- 0.0371$ will provide a reasonable match to observations. To see how our models compare to this range, in Fig. \ref{fig:qhe2_qh1} we show the time evolution of $Q(\HeII)/Q(\HI)$ for two rotation rates $v/v_{\rm{crit}}=0.0$ (top panel) and $0.4$ (bottom panel), and three metallicities $[\mathrm{Fe}/\mathrm{H}] = 0.0$, $-1.0$, and $-2.0$ (blue solid, orange dashed, and green dotted lines respectively), again focusing in the case of highly clumped winds. We indicate our target range of $Q(\HeII)/Q(\HI)$ by the blue band.

We see that all three metallicites can produce ratio in this range during the period immediately after the most massive stars enter the WR phase at $\sim 3$ Myr, regardless of rotation rate. However, for non-rotating stars, only the two higher-metallicity cases remain in this range for $\approx 1-1.5$ Myr, while the lowest metallicity case enters the range only briefly. On the other hand, for the rotating case, all three metallicities remain in this range for $\sim 1.5 -2.5$ Myr. Nonetheless, Fig. \ref{fig:qhe2_qh1} clearly demonstrates that populations of single massive stars with very clumpy atmospheres can produce $Q(\HeII)/Q(\HI)$ ratios that are in principle high enough to reproduce the $I(\HeII)/I({\rm{H}}\beta)$ ratios found in \HeII-emitting nebulae.

\subsection{Nebular spectral synthesis}\label{ssec:syn_spec_sec}

While Sect. \ref{sssec:qratio_sec} is suggestive, our calculations with \textsc{cloudy}, as detailed in Sect. \ref{sec:specmod_sec}, provide more realistic results that include all the complexity we neglected in our simple analytic approximation. They also enable us to examine other observables such the equivalent widths of the \HeII~and blue bump lines. To this end, we generated synthetic nebular spectra using the procedure outlined in Sect. \ref{sec:specmod_sec} at stellar population ages of $2.0$, $2.8$, $2.9$, $3.0$, $3.1$, $3.2$, $3.3$, $3.5$, $3.8$, $4.0$, $4.5$, and $5.0$ Myr. These times provided good coverage of the range of times over which Fig. \ref{fig:qhe2_qh1} suggests that stellar populations are in the right range of ionizing photon ratio to produce line emission consistent with observations.

\subsubsection{I(HeII)/I(H$\beta$)}\label{sssec:iheii_ihbeta_sec}
%%%%%%%%%%%%%%%%%%%%%%%%%%%%%%%%%%%%%%%%%%%%%%%%%%%%%%%%%%%%%%%%%%%%%%%%%%%%%%%%%%%%%%%
\begin{figure}
   \centering
   \includegraphics[width=1.0\columnwidth]{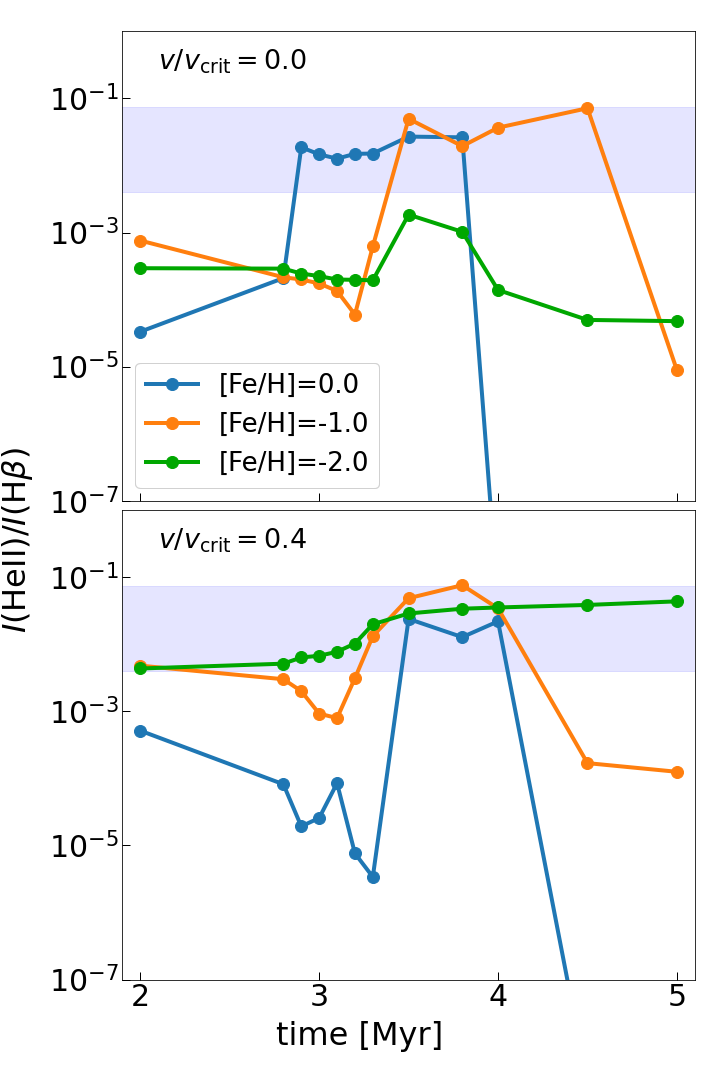}
      \caption{Time evolution of $I(\HeII)/I({\rm{H}}\beta)$ for stellar populations with highly clumped winds at three metallicities, $[\mathrm{Fe}/\mathrm{H}] = 0.0$ (blue), $-1.0$ (orange), and $-2.0$ (green), and two rotation rates, $v/v_{\rm{crit}} =0.0$ (top panel) and $0.4$ (bottom panel).}
         \label{fig:IHe2_IH1}
\end{figure}
%%%%%%%%%%%%%%%%%%%%%%%%%%%%%%%%%%%%%%%%%%%%%%%%%%%%%%%%%%%%%%%%%%%%%%%%%%%%%%%%%%%%%%%

In Fig. \ref{fig:IHe2_IH1} we show the time evolution of the line intensity ratio $I(\HeII)/I({\rm{H}}\beta)$ for three metallicities, $[\mathrm{Fe}/\mathrm{H}] = 0.0$ (blue), $-1.0$ (orange), and $-2.0$ (green), and two rotation rates, $v/v_{\rm{crit}} =0.0$ (the top panel) and $0.4$ (the bottom panel). We show the observed values of $I(\HeII)/I({\rm{H}}\beta) \sim 0.004 \-- 0.07$  for the \HeII\, nebulae in the Cartwheel galaxy \citep{mayya2023} by the blue shaded region. The results are generally consistent with what we might have expected from Fig. \ref{fig:qhe2_qh1}: both the $[\mathrm{Fe}/\mathrm{H}] = 0.0$ and $-1.0$ models go through $\sim 1\-- 1.5\, \mathrm{Myr}$-long intervals where the line intensity ratio is in the observed band, independent of the rotation rate or gas density, while $[\mathrm{Fe}/\mathrm{H}] = -2.0$ remains in this range for a much longer duration only for rapid rotation and is well below it for zero rotation. For all of these models, the maximum of $I(\HeII)/I({\rm{H}}\beta)$ occurs at the time at which the most massive stars ($\gtrsim 70 \, \mathrm{M}_\odot$) are in their WR phases at stellar ages of $\sim 3-4$ Myr. 

We can understand the physical origins of several features of these plots. First, for the non-rotating case we notice that the $[\mathrm{Fe}/\mathrm{H}] =-1$ case yields the highest $I(\HeII)/I({\rm{H}}\beta)$ ratios and the longest duration within the $\sim 0.004 \-- 0.07$ characteristic of observed \HeII-emitting nebulae. This occurs because at $[\mathrm{Fe}/\mathrm{H}]=-1$ represents the optimal compromise between two effects: at high metallicity the surface opacity is too high to produce many \HeII-ionizing photons, while at low metallcity the mass-loss rate is too low to expose the stellar core and produce high $T_\mathrm{eff}$; the $[\mathrm{Fe}/\mathrm{H}]=-1$ case balances these two effects, producing a surface temperature that is enough and a surface opacity that is low enough to maximize \HeII-ionizing photon production. For rapidly rotating stars, by contrast, high metallicity is not needed to produce high mass loss due to the rotation enhanced mass-loss rates and rotational mixing described above; consequently, for the rapidly rotating case, we find that time spent at high $I(\HeII)/I({\rm{H}}\beta)$ increases monotonically as metallicity decreases.

\subsubsection{Line equivalent widths}\label{sssec:ew_sec}
%%%%%%%%%%%%%%%%%%%%%%%%%%%%%%%%%%%%%%%%%%%%%%%%%%%%%%%%%%%%%%%%%%%%%%%%%%%%%%%%%%%%%%%
\begin{figure}
   \centering
   \includegraphics[width=1.0\columnwidth]{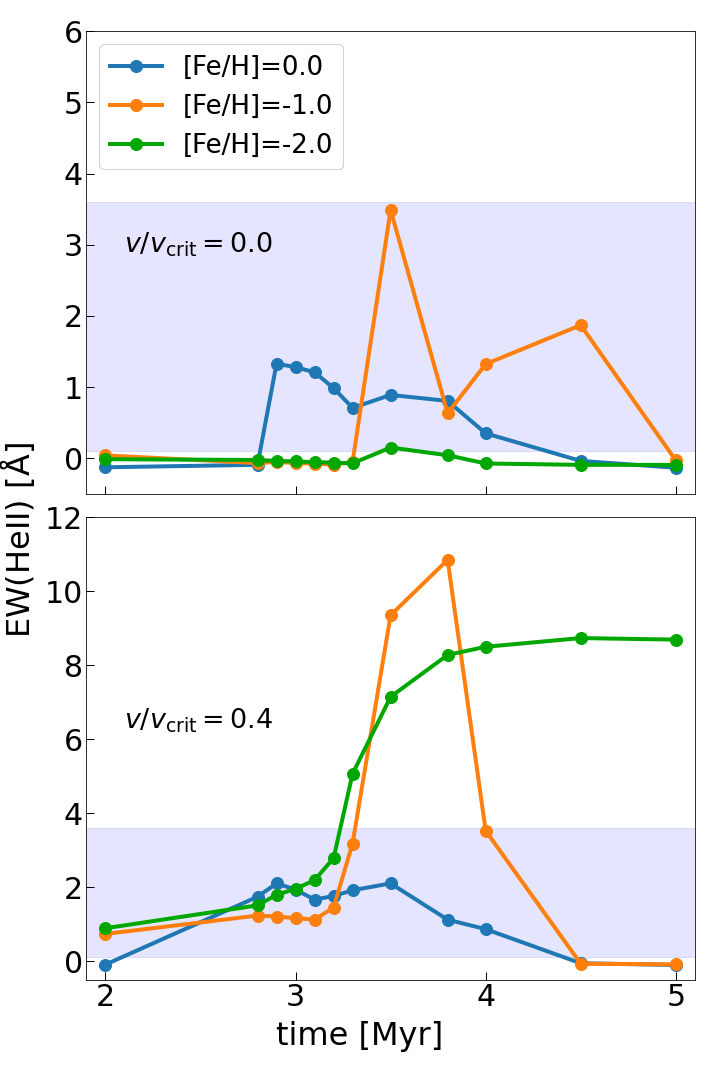}
      \caption{Same as Fig. \ref{fig:IHe2_IH1} but now showing \HeII~EW.}
         \label{fig:ew_he2}
\end{figure}
%%%%%%%%%%%%%%%%%%%%%%%%%%%%%%%%%%%%%%%%%%%%%%%%%%%%%%%%%%%%%%%%%%%%%%%%%%%%%%%%%%%%%%%

%%%%%%%%%%%%%%%%%%%%%%%%%%%%%%%%%%%%%%%%%%%%%%%%%%%%%%%%%%%%%%%%%%%%%%%%%%%%%%%%%%%%%%%
\begin{figure}
   \centering
   \includegraphics[width=1.0\columnwidth]{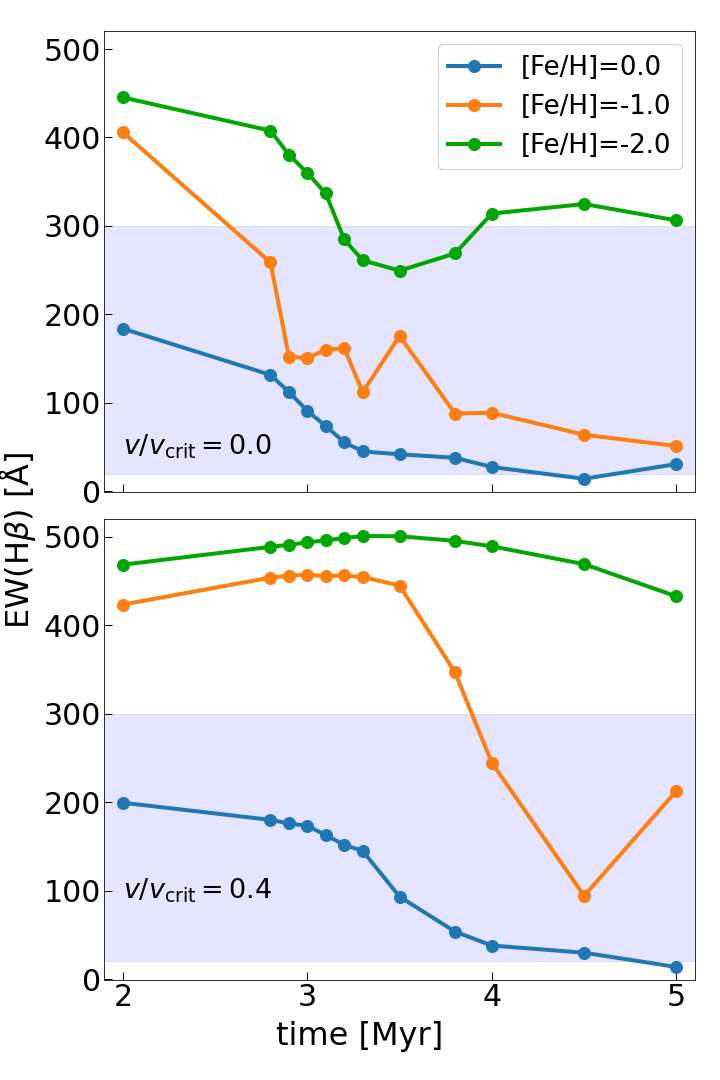}
      \caption{Same as Fig. \ref{fig:ew_he2} but for H$\beta$.}
         \label{fig:ew_hbeta}
\end{figure}
%%%%%%%%%%%%%%%%%%%%%%%%%%%%%%%%%%%%%%%%%%%%%%%%%%%%%%%%%%%%%%%%%%%%%%%%%%%%%%%%%%%%%%%

%%%%%%%%%%%%%%%%%%%%%%%%%%%%%%%%%%%%%%%%%%%%%%%%%%%%%%%%%%%%%%%%%%%%%%%%%%%%%%%%%%%%%%%
\begin{figure}
   \centering
   \includegraphics[width=1.0\columnwidth]{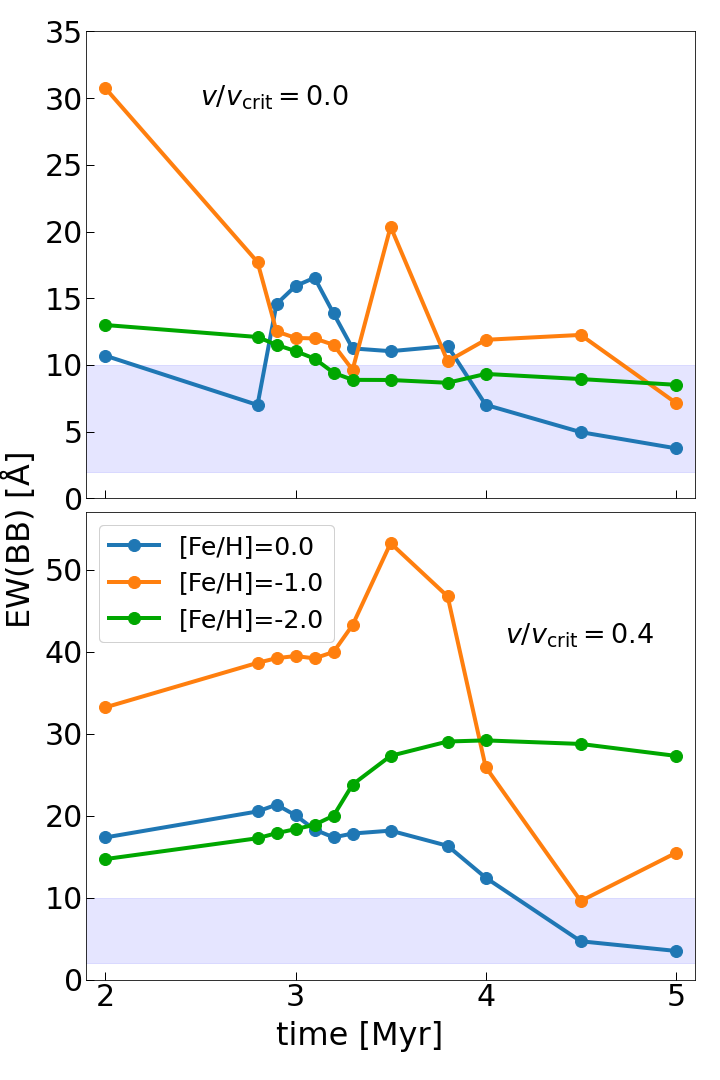}
      \caption{Same as Figs. \ref{fig:ew_he2} and \ref{fig:ew_hbeta} but for the blue bump line blend feature.}
         \label{fig:ew_bb}
\end{figure}
%%%%%%%%%%%%%%%%%%%%%%%%%%%%%%%%%%%%%%%%%%%%%%%%%%%%%%%%%%%%%%%%%%%%%%%%%%%%%%%%%%%%%%%

In addition to examining line ratios, we can also check that our clumpy wind models yield line equivalent widths (EWs) that are consistent with observations. We specifically check the widths of the \HeII, H$\beta$, and blue bump lines. To calculate the EWs of the \HeII\, and H$\beta$ lines, we determine the line fluxes by integrating the nebular plus stellar spectra output by \textsc{cloudy} over $40 \, \angstrom$-wide bands centered at $4685.5$ and $4865 \, \angstrom$ respectively. We obtain the corresponding continuum fluxes for these two lines by integrating over regions on the either side of the line of interest chosen to contain no or only very weak lines; for the \HeII\, line we obtain the continuum level by integrating the spectra from $4760 \-- 4840\, \angstrom$ and from $4540 \-- 4620\, \angstrom$ and averaging the two results, while for  H$\beta$ line we integrate from $4885 \-- 4935 \, \angstrom$ and $4790 \-- 4840 \, \angstrom$ and average the results. We calculate the blue bump line flux by integrating the spectra from $4570 \-- 4740 \, \angstrom$, and the corresponding continuum flux as the average of results obtained by integrating the spectrum over $50\, \angstrom$ windows centered at $4525 \, \angstrom$ and $4905 \, \angstrom$, corresponding to the red and blue sides of the blue bump feature; our procedure here follows that of \citet{mayya2023}.  

We show the time evolution of the EWs for our three lines of interest in Fig. \ref{fig:ew_he2}, Fig. \ref{fig:ew_hbeta}, and Fig. \ref{fig:ew_bb}, respectively. Examining these figures, we see that during the epochs from $\approx 3-4$ Myr when our models predict $I(\HeII)/I(\mathrm{H}\beta)$ ratios sufficiently large to match observations, the typical EWs of the \HeII $\lambda 4686$, H$\beta$, and blue bump lines in the range $\sim 0.1-3.5\,\angstrom$, $\sim 20-300\,\angstrom$, and $\sim 2-10\,\angstrom$, respectively. Exact values vary with metallicity and rotation rate, but these values are in good agreement with the typical ranges found in \HeII-emitting regions by \citet{mayya2020, mayya2023}. We therefore conclude that not only do clumpy wind models for single stars produce $I(\HeII)/I(\mathrm{H}\beta)$ ratios similar to observed values, they also reproduce the observed line-to-continuum ratios in these two lines as well as in the blue bump feature.

\subsubsection{Dependence on wind clumpiness}\label{sssec:clumpywind_subsubsec}
%%%%%%%%%%%%%%%%%%%%%%%%%%%%%%%%%%%%%%%%%%%%%%%%%%%%%%%%%%%%%%%%%%%%%%%%%%%%%%%%%%%%%%%
\begin{figure*}
   \centering
   \includegraphics[width=0.95\textwidth]{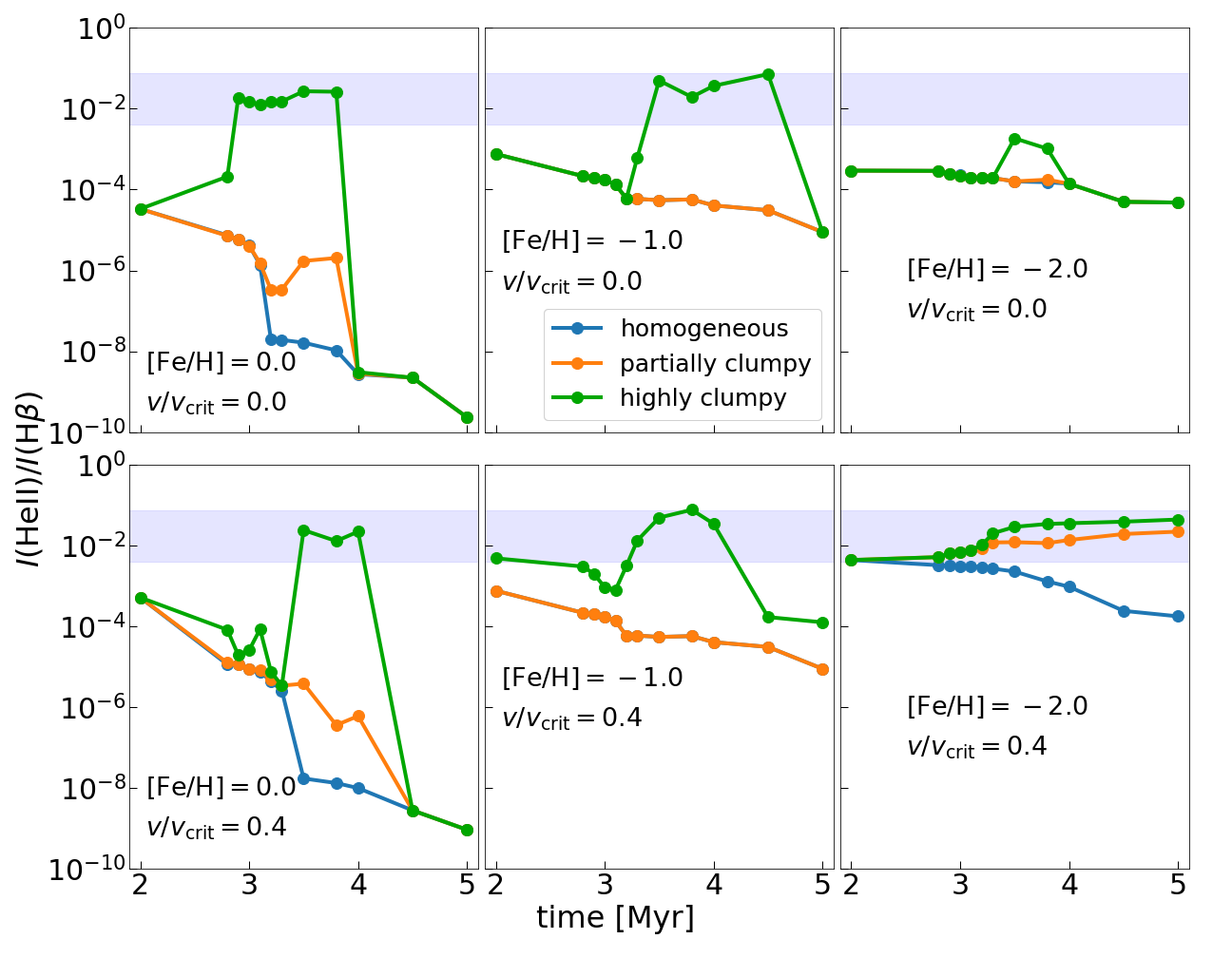}
      \caption{Same as Fig. \ref{fig:IHe2_IH1} except here different colors corresponds to different wind morphologies, ranging from optically thick and homogeneous winds (blue) to partially clumpy (orange) to highly clumpy, porous, and on the verge of optically thin winds (green). Each panel shows a different combination of metallicity and stellar rotation rate: columns show, from left to right, $\mathrm{[Fe/H]} = 0$, $-1$, $-2$; the rows show $v/v_\mathrm{crit} = 0$ (top) and 0.4 (bottom).}
         \label{fig:IHe2_IH1_Dinf_comp}
\end{figure*}
%%%%%%%%%%%%%%%%%%%%%%%%%%%%%%%%%%%%%%%%%%%%%%%%%%%%%%%%%%%%%%%%%%%%%%%%%%%%%%%%%%%%%%%

%%%%%%%%%%%%%%%%%%%%%%%%%%%%%%%%%%%%%%%%%%%%%%%%%%%%%%%%%%%%%%%%%%%%%%%%%%%%%%%%%%%%%%%
\begin{figure*}
   \centering
   \includegraphics[width=0.95\textwidth]{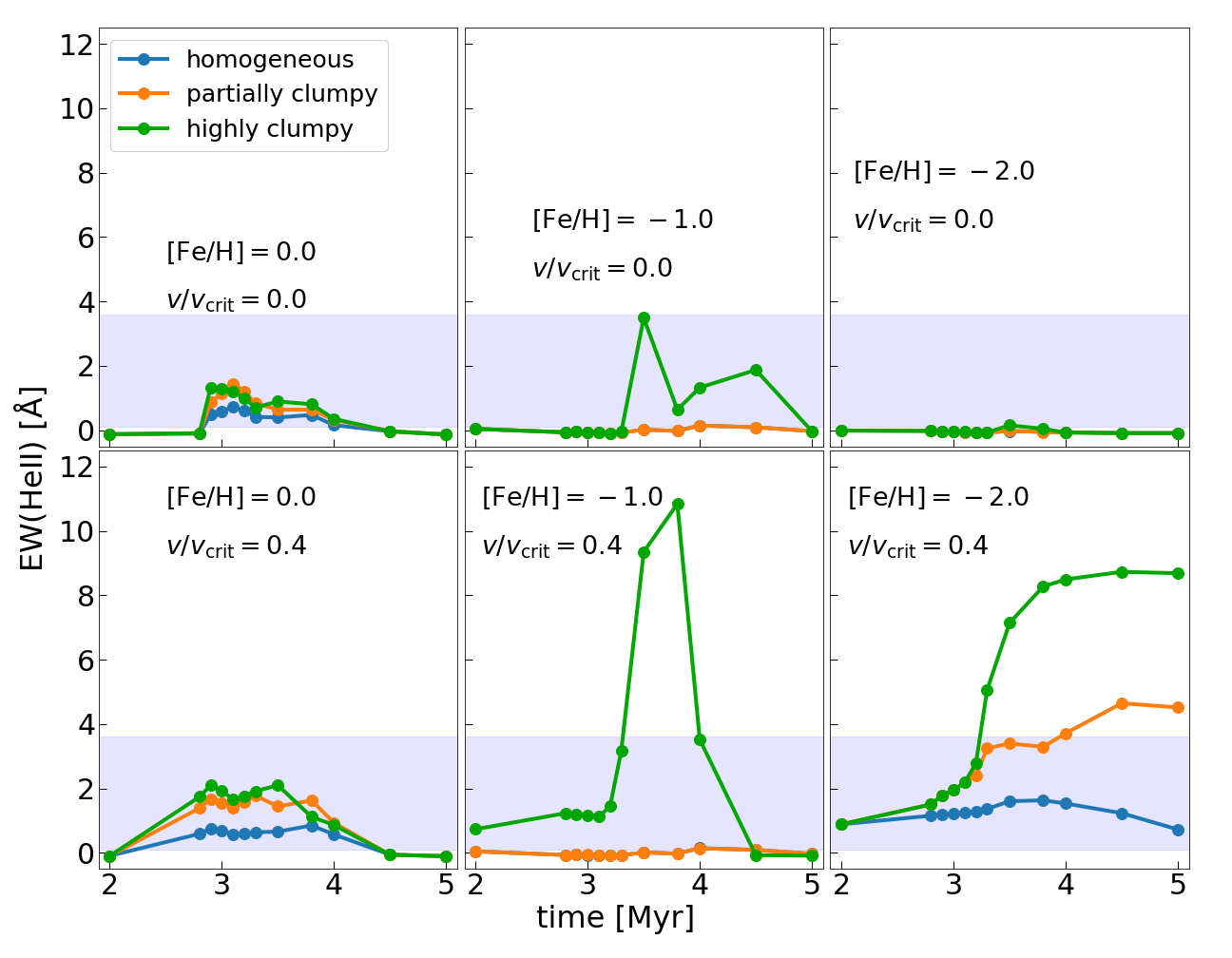}
      \caption{Same as Fig. \ref{fig:ew_he2} except that here we show a comparison for the three atmosphere morphologies. The three metallicity cases are in three different panels: $[\mathrm{Fe}/\mathrm{H}]=0.0$ (leftmost panels), $-1.0$ (middle panels), and $-2.0$ (rightmost panels).} 
         \label{fig:ew_he2_Dinf_comp}
\end{figure*}
%%%%%%%%%%%%%%%%%%%%%%%%%%%%%%%%%%%%%%%%%%%%%%%%%%%%%%%%%%%%%%%%%%%%%%%%%%%%%%%%%%%%%%%

Thus far we have focused on our highest-clumping case, demonstrating that it yields line ratios and EWs that are in very good agreement with observations. We now return to our other two cases, no clumping, smooth winds and intermediate clumping, using our full pipeline to predict $I(\HeII)/I({\rm{H}}\beta)$ ratios and EW for the \HeII\, line exactly as we have for the highly clumped case. We show the results in Fig. \ref{fig:IHe2_IH1_Dinf_comp} and Fig. \ref{fig:ew_he2_Dinf_comp}, respectively.

We find that for both solar and 1/10th solar metallicity, only the highly clumped and porous winds, which is on the verge of the optically thin scenario, can produce the observed $I(\HeII)/I({\rm{H}}\beta)$ ratio $\sim 0.004 \-- 0.07$. However, for the most metal-poor case, $[\mathrm{Fe}/\mathrm{H}]=-2.0$ and rapidly rotating stars, $v/v_\mathrm{crit} = 0.4$, even partially clumpy winds reproduce the observed $I(\HeII)/I({\rm{H}}\beta)$ ratio. This case, while it has a lower \HeII~EW than the highly clumped case, nonetheless reaches EWs of $\approx 4~\angstrom$ at the $\approx 3-5$ Myr ages when the $I(\HeII)/I({\rm{H}}\beta)$ ratio is high, and thus it is well within the observed range for \HeII emitters. We therefore conclude that the combination of efficient rotation-driven CHE (for stars $\gtrsim 90\, \mathrm{M}_\odot$), mass loss, and low atmospheric opacity in this case is sufficient to produce enough \HeII-ionizing photons to match observations even if the stellar surface is partly blocked by a semi-opaque wind.

\section{Discussion and conclusions}\label{sec:conclusion_sec}

The origin of nebular \HeII\, emission in SF galaxies remains an unresolved problem. The line is powered by the recombination of He$^{++}$, production of which requires photons with energies $\geq 54.4$ eV, which are thought to be difficult to produce with conventional single stellar populations. The challenge of explaining the production site of the required photons has led to a number of proposals, including photoionization of He$^+$ by binary stellar populations or by compact objects such as HMXBs and ULXs, shock-driven ionization, and hidden AGNs. By contrast, ionization by single massive and/or WR stars has often been dismissed as a possibility because their \HeII\, ionizing photon production rates were thought to be too low.

However, the stellar models that were the basis for this conclusion have a number of limitations. They treated stellar winds as spherically homogeneous and optically thick, which is at odds with modern observational constraints. They assumed that the responsible stars were non-rotating, contrary to theoretical expectations that massive stars should be born with high rotation rates \citep{rosen2012}. In this paper, we removed these limitations and revisited the single WR channel as a possible source for the nebular \HeII\, emission using a dense grid of massive stellar evolution models at a range of metallicities and rotation rates coupled to three different WR wind morphologies. We summarize our primary findings in the following:

\begin{itemize}
\item We find that varying our treatment of clumping in extended WR atmospheres leads to dramatic variations in the predicted \HeII-ionizing photon budgets. Models that assume highly clumped winds that expose most of the stellar surface yield a factor of 20 increase in the production of photons beyond the He$^+$ ionization edge compared to traditional models, where winds are treated as spherically homogeneous and optically thick. This enhanced production of hard photons occurs for stellar populations with ages of $\approx 3-5$ Myr, corresponding to the time at which the most massive stars, $\gtrsim 70$ M$_\odot$, enter the WR phase for $[\mathrm{Fe}/\mathrm{H}]\gtrsim -2.0$.
\item During this period of enhanced hard photon production, nebular modeling with \textsc{cloudy} revealed that nebulae powered by stellar populations with highly clumped winds experience $\sim 1-2$ Myr periods where their ratio of \HeII~$\lambda4686$ to H$\beta$ line luminosities reaches the large values, $I(\HeII)/I(\mathrm{H}\beta) \sim 0.004 - 0.07$, observed in local \HeII-emitting regions. During this period, the equivalent widths of the \HeII, H$\beta$, and blue bump lines are also comparable to the observed values -- a few Angstroms, a few hundred Angstroms, and a few tens of Angstroms, respectively. 
\item The periods where the predicted nebular emission closely matches what we observe occur for all metallicities and stellar rotation rates if winds are highly clumped. If we treat WR winds as moderately clumped and elevated $I(\HeII)/I(\mathrm{H}\beta)$ ratios and \HeII~equivalent widths still occur for metal-poor rapidly rotating stellar populations ($[\mathrm{Fe}/\mathrm{H}]= -2$, $v/v_\mathrm{crit} = 0.4$) but not for more metal-poor or non-rotating ones, for a conventional model using smoothed winds, there are no periods of elevated $I(\HeII)/I(\mathrm{H}\beta)$ for any of the metallicities or rotation rates we have considered, consistent with earlier work.
\end{itemize}

We therefore conclude that earlier estimates that single stars with moderate rotation rates, $v/v_\mathrm{crit} \lesssim 0.5$, could not be the drivers of nebular \HeII~emission are incorrect since this conclusion holds only if we assume that these stars have spherically homogeneous and non-clumpy winds. On the contrary, single WR stars with highly clumpy to partially clumpy winds are strong candidates for drivers of the nebular \HeII\, emission seen in both local and high-redshift galaxies.

Our findings have important implications, particularly at high redshifts. Massive stars in metal-poor environments are typically expected to be moderately to rapidly rotating, with $v/v_{\rm{crit}} \gtrsim 0.4$ \citep{rosen2012}, and high-redshift galaxies may also have IMFs that are more abundant in massive stars \citep[e.g.,][]{sharda2022, van-Dokkum24a}. The combination of these factors likely results in a greater abundance of massive rotating stars in the early Universe, facilitating the contribution of single massive stars as significant sources of \HeII\, emission. This may explain the increasing abundance of \HeII emitters in the early Universe and may have further implications for \HeII\, reionization.

\begin{acknowledgements}
Authors thank the anonymous referee for the useful suggestions to improve the paper. Authors also thank Daniel Schaerer and Anastasios Fragkos for the insightful discussions on the current challenges of nebular \HeII\, emission. Moreover, authors thank Andrea Ferrara for extremely useful discussions on the \HeII\, emissions at high-redshift galaxies. AR gratefully acknowledges the support from Andrea Ferrara's Italian funding scheme ``The quest for the first stars” (Cod. 2017T4ARJ5\_001). SS and AP also acknowledge the support from the Italian funding scheme ``The quest for the first stars” (Cod. 2017T4ARJ5\_001). MRK acknowledges support from the Australian Research Council through its ``Laureate Fellowship" scheme, award FL220100020. GM and SE have received funding from the European Research Council (ERC) under the European Union's Horizon 2020 research and innovation programme (grant agreement No 833925, project STAREX). JSV acknowledges support from STFC (Science and Technology Facilities Council) funding under grant number ST/V000233/1. AS is supported by the German Deutsche Forschungsgemeinschaft (DFG) under Project-ID 445674056 (Emmy Noether Research Group SA4064/1-1, PI Sander).The simulation data for this research/project were produced with the assistance of resources and services from the Scuola Normale Superiore's Center for High Performance Computing (CHPC), and the National Computational Infrastructure (NCI)’s supercomputer Gadi, supported by the Australian Government (award jh2). Additional software used include \texttt{SciPy} \citep{virtanen2020}, \texttt{matplotlib} \citep{hunter2007}, and \texttt{NumPy} \citep{harris2020}.
\end{acknowledgements}

\section*{Data availability}
The data underlying this article will be shared on request to the corresponding author (AR).

\bibliographystyle{aa} % style aa.bst
\bibliography{refs.bib}

\begin{appendix} 

\section{On the importance of mass resolution}\label{app:massres}

As discussed in the main text, we use MIST tracks in part because of the high mass resolution they offer at the high mass end of the IMF. To demonstrate why this is important, we compare results generated using the high mass-resolution that is the default for MIST to results generated using the same MIST tracks, but with the mass-resolution reduced to match that of the Geneva stellar tracks\footnote{\href{https://www.unige.ch/sciences/astro/evolution/en/database/}{https://www.unige.ch/sciences/astro/evolution/en/database/}} \citep{ekstrom2012, georgy2013}, which are the most widely used for massive stars but have much lower mass resolution at the upper end of the mass range; specifically, they include models at initial mass $M = 60, 85$ and $120$ M$_\odot$ only.

%%%%%%%%%%%%%%%%%%%%%%%%%%%%%%%%%%%%%%%%%%%%%%%%%%%%%%%%%%%%%%%%%%%%%%%%%%%%%%%%%%%%%%%
\begin{figure}
   \centering
   \includegraphics[width=1.0\columnwidth]{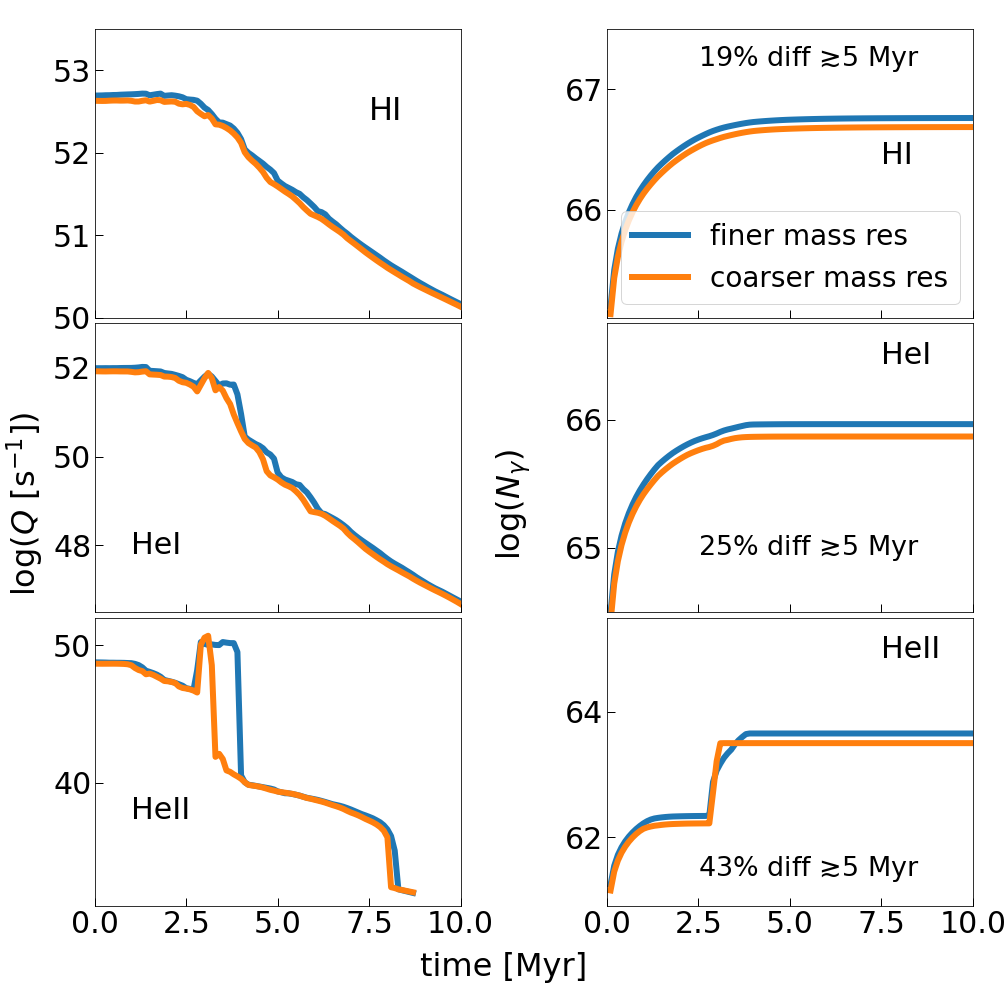}
      \caption{Same as Fig. \ref{fig:atm_comp} but now comparing the results generated using MIST models with finer (blue) and coarser (orange) mass resolutions. For both sets of models we show the case of highly clumped, close to optically thin winds.}
         \label{fig:massRes}
\end{figure}
%%%%%%%%%%%%%%%%%%%%%%%%%%%%%%%%%%%%%%%%%%%%%%%%%%%%%%%%%%%%%%%%%%%%%%%%%%%%%%%%%%%%%%%

To demonstrate the importance of resolution, in Fig. \ref{fig:massRes} we show the time evolution of the \HI\,, \HeI\,, and \HeII\, ionizing luminosities for highly clumped, optically thin atmospheres computed using MIST-I high and low resolution stellar tracks, holding all other parameters at the same values used in Fig. \ref{fig:atm_comp}. The blue line shown here for the MIST-I models is identical to the highly clumped case shown in Fig. \ref{fig:atm_comp}. For the low-resolution tracks we observe kinks in the \HeI\, and \HeII\, ionizing luminosities as stars $\gtrsim 70 \, \mathrm{M}_\odot$ enter the WR phase, similar to the MIST results, but we find that this peak is substantially broader in time for the finer mass-resolution compared to the coarser one, resulting in cumulative differences of approximately 25\% and 45\% for \HeI\, and \HeII\, respectively, with a comparatively smaller difference in \HI\, luminosities of $\sim 20\%$ at $\gtrsim 5$ Myr. The difference between these two mass-resolution cases is simply a result of the need for far more interpolation across the mass range that produces WR stars in the lower-resolution tracks than for the higher-resolution ones.

\section{Dependence on initial abundance}\label{app:trackcomp}
\begin{figure}
   \centering
   \includegraphics[width=1.0\columnwidth]{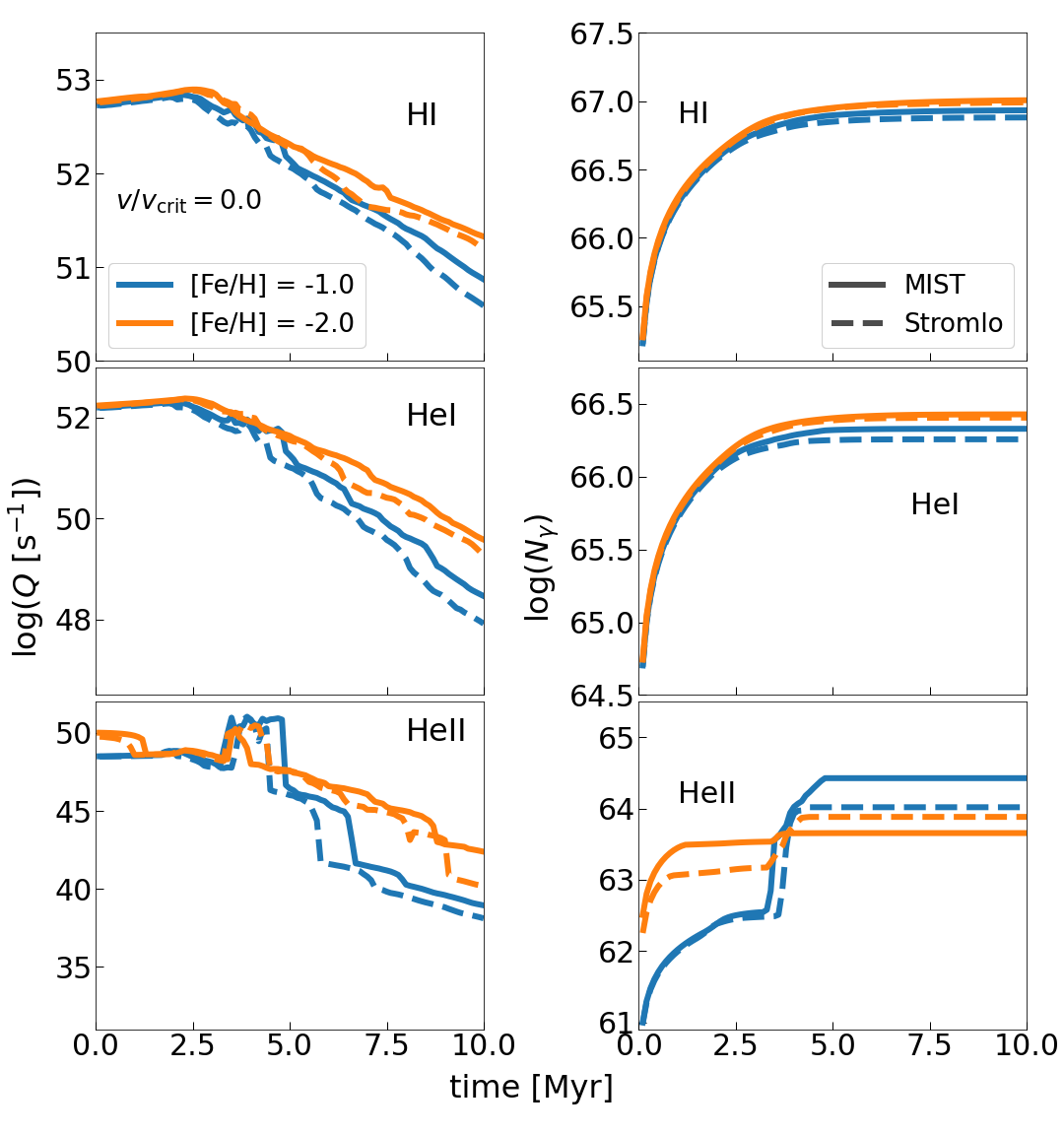}
      \caption{Same as Fig. \ref{fig:atm_comp} but now comparing the results generated using MIST and Stromlo stellar tracks to show the dependence on initial metal  abundances at a given metallicity. The tracks shown here are for two metallicities, $[\mathrm{Fe}/\mathrm{H}]=-1$ and $-2$, and for non-rotating stars with highly clumped, close to optically thin winds.}
         \label{fig:tracks}
         
\end{figure}

As discussed in Sect. \ref{sec:stelmod_sec}, for the main part of the paper we use MIST tracks, which assume solar-scaled abundances. However, we find in Sect. \ref{sssec:feh_sec} that metallicity plays a crucial role in determining mass-loss rates, and consequently, the ionizing photon budgets, particularly for non-rotating stars, and it is therefore important to verify that these results do not change significantly if we adopt more realistic non-solar-scaled abundances. For this purpose we compare the results from MIST to results generated using the Stromlo Stellar Tracks (SST; \citealt{grasha2021})\footnote{Note that both MIST and SST have exactly same physics models and parameters, the only difference comes in their initial abundance setups.}, focusing only on non-rotating tracks, since we have seen in Sect. \ref{sssec:feh_sec} that surface parameters are independent of metallicities for stars with $v/v_{\rm{crit}}=0.4$, and on metallicities $[\mathrm{Fe}/\mathrm{H}] = -1$ and $-2$, since MIST and SST are nearly identical at solar metallicity. We show the time evolution of the instantaneous and cumulative ionizing photon outputs for these cases in Fig. \ref{fig:tracks}.

%%%%%%%%%%%%%%%%%%%%%%%%%%%%%%%%%%%%%%%%%%%%%%%%%%%%%%%%%%%%%%%%%%%%%%%%%%%%%%%%%%%%%%%
%%%%%%%%%%%%%%%%%%%%%%%%%%%%%%%%%%%%%%%%%%%%%%%%%%%%%%%%%%%%%%%%%%%%%%%%%%%%%%%%%%%%%%%

For softer photons (\HI\, and \HeI), we see that MIST and SST produce similar results, with maximal percentage differences in cumulative photon luminosities of $\sim 12.8\%$ and $\sim 3.7\%$ for \HI\,, and $\sim 18\%$ and $\sim 5\%$ for \HeI\, at $[\mathrm{Fe}/\mathrm{H}] = -1$ and $-2.0$, respectively. However, the disparity increases for harder photons (\HeII), where the maximal differences reach $\sim 155\%$ and $\sim 68\%$ for $[\mathrm{Fe}/\mathrm{H}] = -1$ and $-2$, respectively. While this difference is not negligible, it is also not as pronounced as those induced by other parameter choices, and in particular is much smaller than the effects of our choice of atmosphere model. We therefore conclude that our results remain robust regardless of the initial abundance distribution, and adopt the solar-scaled abundance pattern throughout the main text.

\end{appendix}

\end{document}